\documentclass[runningheads]{llncs}
\usepackage{subcaption}
\usepackage[labelformat=parens,labelsep=quad,skip=3pt]{caption}

\usepackage{graphicx}

\usepackage{subcaption}
\usepackage[labelformat=parens,labelsep=quad,skip=3pt]{caption}
\usepackage{booktabs}
\usepackage{varwidth}
\usepackage{graphicx}
\usepackage{url}

\newcommand{\mrmralgo}{mRMRe} 

\newcommand{\ra}{\it RA}  
\newcommand{\rl}{\it RL}   
\newcommand{\sd}{\it SD} 
\newcommand{\fs}{\it FS} 
\newcommand{\rct}{\it RCT} 

\newcommand{\FRA}{Feature\,\, Range\,\,  Analysis} 
\newcommand{\RA}{Range\,\,  Analysis}  
\newcommand{\RL}{Rule \,\,  Learning}

\newcommand{\SD}{Subgroup \,\, Discovery}

\newcommand{\FS}{Feature \,\, Selection}
\newcommand{\MRMR}{MRMR}

\newcommand{\TP}[1]{TP({#1})}
\newcommand{\FP}[1]{FP({#1})}
\newcommand{\TN}[1]{TN({#1})}
\newcommand{\FN}[1]{FN({#1})}
\newcommand{\TPR}[1]{TPR({#1})}
\newcommand{\FPR}[1]{FPR({#1})}

\newcommand{\PPV}[1]{PPV({#1})} 
\newcommand{\FOS}[1]{F1Score({#1})}
\newcommand{\Lift}[1]{Lift({#1})}

\newcommand{\NPLR}[1]{NPLR({#1})}
\newcommand{\WRAcc}[1]{WRAcc({#1})}
\newcommand{\ROC}[1]{ROCAcc({#1})}
\newcommand{\Acc}[1]{Acc({#1})}  
\newcommand{\CWP}[1]{CW\!P({#1})} 

\newcommand{\AllIn}[1]{n({#1})} 
\newcommand{\All}{{\cal N}} 
\newcommand{\Pos}{Pos} 

\newcommand{\feat}{{\cal F}}  

\newcommand{\level}{l}  
\newcommand{\rangef}{{\cal R}}  

\newcommand{\pairf}{{\cal P}} 

\newcommand{\rpafeat}{nRanking} 
\newcommand{\rpamrmr}{nBasis} 
\newcommand{\sample}{{s}} 

\newcommand{\occ}[1]{Occurs({#1})} 
\newcommand{\cnt}[1]{Count({#1})} 
\newcommand{\ooo}[2]{OOO({#1}, {#2})} 
\newcommand{\ino}[2]{INO({#1}, {#2})} 
\newcommand{\timeo}[1]{Timing({#1})} 
\newcommand{\timet}[2]{Timing({#1}, {#2})} 
\newcommand{\pc}{PkgC8} 

\newcommand{\histo}[1]{H({#1})} 
\newcommand{\distr}[1]{D({#1})} 
\newcommand{\events}[1]{E({#1})} 

\newcommand{\delete}[1]{} 
\newcommand{\todo}[1]{}
\newcommand{\qa}[1]{}

\begin{document}

\title{Accelerating System-Level Debug Using Rule Learning and Subgroup Discovery Techniques}
\titlerunning{Accelerating System-Level Debug Using RL and SD Techniques}
%

\author{Zurab Khasidashvili} 
\authorrunning{Z. Khasidashvili}
\institute{Intel Israel Development Center, Haifa, Israel\\
\email{zurab.khasidashvili@intel.com}}

\maketitle              
%

\begin{abstract}
We propose a root-causing procedure for accelerating system-level debug using rule-based techniques. We describe the procedure and how it provides high quality debug hints for reducing the debug effort. This includes heuristics for engineering features from logs of many tests, and the data analytics techniques for generating powerful debug hints. As a case study, we used these techniques for root-causing failures of the Power Management design feature Package-C8 and showed their effectiveness. Furthermore, we propose an approach for mining the root-causing experience and results for reuse, to accelerate future debug activities and reduce dependency on validation experts. We believe that these techniques are beneficial also for other validation activities at different levels of abstraction, for complex hardware, software and firmware systems, both pre-silicon and post-silicon.
\end{abstract}

\section{Introduction}

An estimated $50-60\%$ of validation effort is spent on debugging. Debugging in a system-level platform represents a great challenge due to the enormous number of involved software, firmware and hardware components, domains, flows, protocols, design features, interfaces etc. The current practice is to follow the traces of design failures as they cross the system domains and allocate the right experts to assist in debugging sub-flows associated to these domains, starting from the sub-flows where the mismatches from the expected behavior are first observed.  Another challenge is that there is no established way to learn from root-causing of a failure and mine the learning in order to reuse it for future root-causing activities or to automatically root-cause related failure scenarios in the future. 

For such distributed debug activities, terms \emph{debugging} and \emph{root-causing} might mean different things to different validators as well as in different validation contexts:  If a bug (or an unexpected behavior or outcome of a test) is originating from module B, then for a validator responsible for module A root-causing might mean to understand whether the bug is related to module A, and when there is a sufficient evidence that this is not the case, contact a validator of another module where the bug might potentially hide. For the validator of module B, root-causing might mean to understand and confirm that the bug is originating from the hardware or software code of module B. Furthermore, in case of post-silicon validation, the bug or unexpected behavior can be caused by physical phenomena like manufacturing defects or manufacturing process variations, and in that case the validator of module B might require help from a physical designer or  manufacturing expert to root-cause the unexpected behavior. Root-causing in such a case might mean to trace down the failure to a physical device that exposes a rare or a systematic weakness under some operating conditions (temperature, voltage, power, or other). The failure in that case might not be associated to a line in hardware, software, or firmware code, and instead be associated with phenomena which are currently very difficult or impossible to control.

We propose to use  $\SD$ ($\sd$)~\cite{klosgen1996multipattern,wrobel1997algorithm,van2012diverse,atzmueller2015subgroup} and $\RL$ ($\rl$)~\cite{clark1989cn2,furnkranz2012foundations} techniques for automating the root-causing activity and generating valuable debug hints based on data that is collected from logs of many tests.  The aim of  $\sd$ is to identify ``interesting'' subgroups of data that can be described as feature-range combinations; and the aim of  $\rl$ is to build classification models using \emph{rules} with feature-range pairs (as in $\sd$) as the antecedent and the predicted class as the consequent. $\sd$ and $\rl$ do not operate on trace logs directly, and an important part of our approach is how to \emph{engineer features} from logs to capture information that might potentially be useful for root-causing. The generated debug hints provide the system-level validator with some level of independence to debug and narrow down the debug space, and this way reduce the dependency on the experts. On the other hand, it also relieves the experts from the burden of massive debug support and also helps them with accelerating debugging. The suggested method also provides a platform for knowledge sharing of previous debug activities. We demonstrate our method on an important and representative $\pc$ flow in the Power Management (PM) protocol implemented in Intel processors~\cite{IntelCore11thGEN}.

Specifically, we apply $\FRA$, or $\RA$ ($\ra$) for short~\cite{khasidashvili2019range,khasidashvili2021feature}, on data engineered from trace logs,  for root-causing failing executions.  When the data labels are binary, the aim of $\ra$ is to isolate \emph{subgroups} of the positive samples from the negative ones. In the context of this work, positive samples represent failing flow executions and negative samples represent passing ones. $\ra$ is closely related both to $\sd$ and to $\rl$, and  represents a way of unifying the $\sd$ and $\rl$ approaches: Like $\sd$, it aims at identifying feature-range pairs for isolating subgroups of the data with respect to quality functions such as \emph{Lift} and \emph{WRAcc} (see Section~\ref{SS.FRA}), and in addition it uses $\FS$~\cite{guyon2003introduction} algorithms aiming at selecting both features highly correlated with the class labels and features giving a good coverage of the variation in the data~\cite{ding2005minimum}.  In $\ra$ these high quality feature-range pairs are represented as binary features and they can be used as antecedents of prediction rules as in $\rl$, as well as be used as engineered features along with the original features for training classification models; such models can outperform models built from the original features only~\cite{khasidashvili2021feature}.

 In the iterative  procedure of \emph{root-causing tree} construction proposed in Section~\ref{S.ml},  $\ra$ is used in each iteration to identify finer subgroups of failures such that all failures in a subgroup have the same root cause and different subgroups represent different root causes. In the validation domain, the $\rl$ and $\sd$ techniques have been used for improving test coverage and eliminating redundancy~\cite{chen2013simulation,katz2011learning}, but to the best of our knowledge they have not been used for automating root-causing of test failures.

The paper is organized as follows. In Section~\ref{S.pm}, we describe the basics of the Power Management flow. In Section~\ref{S.events}, we define events, flows, and requirements from traces. In Section~\ref{S.features}, we describe how we engineer features from trace logs and the dataset used for inferring root-causing hints. In Section~\ref{S.ml}, we describe our ML approach. Experimental results are presented in Section~\ref{S.results}. In Section~\ref{S.related}, we explain how our work is different from related work on trace based root-causing. We conclude by discussing future work in Section~\ref{S.future}.

\section{Power Management Flows}\label{S.pm}

Power-Management (PM) is a classic topic for system-level protocols to verify. The global PM flows of the Client $CPU$ involve interactions of various internal and external components controlling the power entities across the system, such as Voltage-Regulators, Power-Gates, Clocks, etc.

The Package C-states ($PkgC$) are a set of design features for putting the $CPU$ in various levels of sleep states, while $\pc$ is one of the deeper $PkgC$ states. The user visible information on PM in Intel processors is described in~\cite{IntelCore11thGEN}, \emph{Chapter~1.4}. The processor supports multiple $PkgC$ states: $C0, C2, C3, C6-10$. State $PkgC0$ is the normal operation state of the processor -- no power saving requirements are imposed. From the $PkgC0$ state the processor can only transition to state $PkgC2$ (which is not a power-saving state) from which it then can transition to any of the power-saving $PkgC$ states $C3, C6-C10$: the higher the number of power-saving state, the higher the power saving in that state. 

The processor remains in the $PkgC0$ state if at least one of the \emph{IA Cores} is in state $C0$ or if the platform has not yet granted the processor the permission to go into power-saving states. In general, a $PkgC$ state request is determined by the lowest state of processor \emph{IA Cores}. A $PkgC$ state is automatically resolved by the processor depending on the idle power states of the processor \emph{Core} and status of other platform components.

\begin{figure}[!ht]
\center
\includegraphics[width= 0.7\columnwidth]{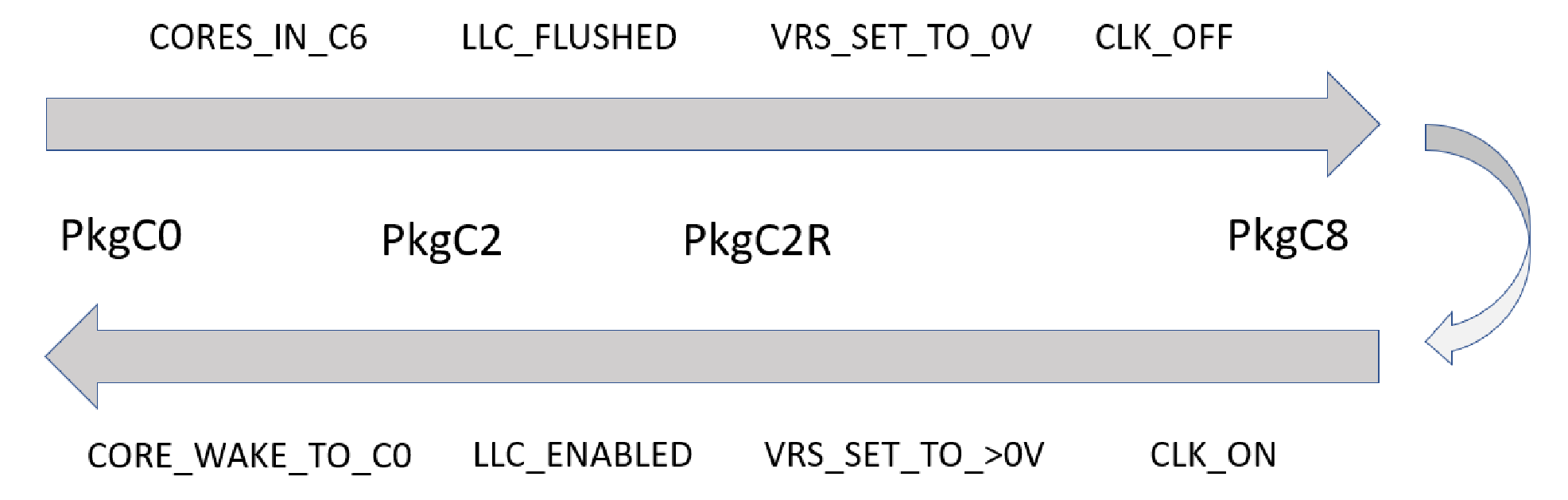}
\caption{PkgC8 entry and exit flows.} \label{PkgC8_flow}
\end{figure}

Figure~\ref{PkgC8_flow} depicts the $\pc$ entry and exit flows in a nutshell; it provides some basic information useful to illustrate the method presented in this paper (familiarity with details of PM flow is not required). This description of the $PkgC$ states does not apply to any specific generation of Intel products and may change in future generations, and therefore it should be treated as an example. Once all \emph{Cores} go into sleep-state, the $PkgC$-entry flow will be triggered, starting with $PkgC2$. If the conditions allow to go deeper, the processor will initiate \emph{Last Level Cache} ($LLC$) flush (reaching $PkgC2R$). The following stages will be setting the \emph{Core}, \emph{Graphics} and \emph{Ring} voltages to $0v$, and turn-off the main \emph{CPU} clock ($Clk$). The $\pc$ wake-up flow will set back the voltages and ungate various components of the \emph{CPU} internal domains.

\section{Representation of Events, Traces and Flows}\label{S.events}

We assume a set of \emph{observable variables} tracked along execution of a test. Variables originating from hardware are often referred to as \emph{signals}.  In a validation environment, an event is usually referred to as a value change of an observable variable. In this work, an \emph{event} is a pair $(variable, value\_range)$ for numeric observable variables or a pair $(variable, level\_subset)$  for categorical observable variables, interpreted as conditions $variable \in value\_range$ and $variable \in level\_subset$, respectively. In case of singleton sets $value\_range = \{v\}$ and $level\_subset = \{l\}$, we write these conditions as $variable = v$ and $variable =~l$. For example, a binary clock signal  $Clk$ is treated as an (un-ordered) categorical variable; two events are associated to it: $CLK\_OFF \equiv Clk = 0$ and $CLK\_ON  \equiv Clk = 1$, which are fired when $Clk$ changes value from $1$ to $0$ or from $0$ to $1$, receptively. These two events cannot occur at the same time but can occur along the same execution. As another example, one way to encode a message $Msg$ with two possible status values  $\{snd, rcv\}$ is to treat it as a categorical variable with these two levels and associate to it two events $MSG\_SND \equiv Msg = snd$ and $MSG\_RCV \equiv Msg = rcv$ that are fired when $Msg$ status changes to sent and received, respectively.

By considering value ranges and level subsets, we allow for an \emph{abstraction} that ignores some of the changes in variable values.  
For ordered categorical variables, grouping adjacent levels with respect to the order into a single level is a meaningful way of abstraction. For numeric variables, there are many clever ways of  \emph{discretization}~\cite{kotsiantis2006discretization} into intervals, both supervised and unsupervised. Because the right discretization for root-causing  is difficult to quickly estimate upfront, our suggestion is to use discretization techniques where selected ranges can overlap and have different lengths (including length zero)~\cite{khasidashvili2019range,khasidashvili2021feature}. Then the root-causing algorithm will have a wider choice to select the most relevant event sets. 

We assume that a \emph{trace} (or \emph{trace log}) is a collection of event \emph{occurrences}, each associated with the following attributes:

\begin{enumerate}
\item  (\emph{Time Stamp}) The \emph{cycle} or a \emph{time stamp} of the event occurrence. Usually, for the purpose of validation or root-causing, the absolute values of time stamps are not relevant and therefore are abstracted away, what is important is that the time information captures the sequential vs concurrency information along a trace. In other contexts, the precise timing information can be critical for detecting anomalous behavior (e.g.,~\cite{stearley2004towards}).
\item  (\emph{Module ID}) The \emph{module} from which the event is originating. This allows one to zoom in into execution of a particular module that contributes to the events of interest, if debugging so far suggests that the root cause might be found in that module. 
\item  (\emph{Instance ID})  An \emph{index} uniquely determining the \emph{instance} of the module from which the event occurrence is originating,  if multiple instances of that module can be executed, including concurrently, along an execution. 
\end{enumerate}

Finally, we assume that a \emph{flow} is succinctly represented as a sequence of events that must occur along the trace of a passing execution (if that flow was triggered). Events that occur in the description of a flow are called \emph{flow-events}. In different languages, flows are specified in different forms such as regular expressions of events or as automata. Work~\cite{fraer2014visual} introduces a very rich visual formalism to specify concurrent multi-agent flows, with well-defined formal semantics. In the Specman~\cite{hollander2001language} environment that we are using, \emph{reference models} can be used to specify flows. The flow formalisms and validation environments define and implement the ability to check the traces against flows and conclude a pass or fail status of the flow execution. In this work we avoid sticking to any particular formalism for flows or introducing a new one.

Many formalisms for analyzing traces of concurrent executions assume that the trace logs satisfy the above requirements (see e.g.~\cite{fraer2014visual,pal2021feature}, Sections V and II, respectively). The logger scripts generating trace logs in validation and test environments that we have experimented with support the above requirements.  This allows one to correctly identify event occurrences associated to a \emph{flow instance} with an instance ID $i$  along the trace, the \emph{trace fragment} associated to that flow instance, and the corresponding pass or fail \emph{labels}. This trace fragment starts with the occurrence of the first event in the flow with ID $i$, and ends with the occurrence of the last event in the flow with ID $i$ in case of a successful execution of the flow, and otherwise ends at the end of the trace or earlier, as soon as criteria allowing to ignore the rest of the trace are met, if any. Here, a successful execution of a flow instance is defined as one where occurrences of all events in the flow with ID $i$ are matched within the trace in the right order, and in that case the associated label is ``pass'', and otherwise the label is ``fail''. 

It is important to note that the events that we use in root-causing are not only the flow-events of the flow whose instances we want to root-cause: Most root-causing approaches try to identify anomalous sub-sequences of a flow, while we want to explain what goes wrong before, in-between, and after the expected flow-events. Our approach to engineering features from traces described in the next section is performed per flow instance and in most cases focuses on event occurrences in the trace fragment associated to it. Note that the above definition of the trace fragment associated to a flow instance, while simple, convenient and sufficient from the validator's point of view for the PM flow root-causing goals, could in general be very restrictive:  the trace prefix leading to the trace fragment associated to the flow instance of interest might contain information critical for root-causing. Actually, also the assumption that future events occurring after the associated trace fragment cannot impact or explain failing executions need not always be valid, for example when execution depends on estimating probabilities of future events based on trends in the data observed so far, or when the effect of an anomalous execution can only be observed late in the execution. In general,  taking into account event occurrences from the entire trace might be beneficial for root-causing, and this can be achieved by appropriately adapting the definition of the trace fragment associated to a flow instance and/or by using the full trace during feature engineering.

\section{Feature Engineering}\label{S.features}

As a preprocessing step of the trace logs, for each test in the regression suite and each instance of the flow of interest in it, we extract the corresponding trace fragment (and will treat it as an independent, full input trace), and associate a pass/fail label to that instance, as described in Section~\ref{S.events}. Each such instance of the flow and the corresponding trace fragment define a row (a sample) in the dataset that we build as input to ML algorithms.

For each event $e$, not necessarily a flow-event, we generate two features: 
\begin{itemize}
\item A binary $1/0$ feature $\occ{e}$ indicating whether event $e$ occurs along the execution of that flow instance.
\item An integer feature $\cnt{e}$ continuing the number of occurrences of event $e$ in the flow instance.
\end{itemize}

Consider an observable variable $v$ with which events $\events{v} =\{e_{vi} \| i \in I\}$ are associated in a flow instance or entire trace, where $e_{vi}=(v, r_i)$ and  $\{r_i \| i \in I\}$ is a  discretization of the full range of $v$ into non-overlapping ranges (bins). Then the counts $\cnt{e_{vi}}$ associated to the events $e_{vi} \in \events{v}$ define a histogram $\histo{v}$ with bins \{$r_i\}$. The distribution $\distr{v}$ approximated by this histogram gives a global view of the behavior of variable $v$ along the trace, and its accuracy is only limited by the granularity of the bins $r_i$ into which the full range of $v$ was discretized. We call $\histo{v}$ and $\distr{v}$ the histogram and the distribution associated to $v$, respectively.

\begin{itemize}
\item We then generate descriptive statistics for that histogram -- min, max, mean, median, std (standard deviation), skewness, kurtosis, IQR (interquartile range), and percentiles -- and add them as new features to the dataset.
\item We also generate a best distribution fit to the histogram from a list of potentially relevant, continuous or discrete distributions, depending on the type of $v$ (uniform, normal, lognormal, Weibull, exponential, gamma, etc.) and add the corresponding categorical feature to the dataset. The coefficients of the best fit for each tried distribution can also be added to the dataset as new features, optionally.
\end{itemize}

In addition, to encode sequential information among the \emph{flow-events}, for each pair of distinct flow-events $(e_1, e_2)$ such that $e_1$ comes before (or optionally, comes immediately before) $e_2$ in the flow, we can generate: 

\begin{itemize}
\item A binary feature $\ino{e_1}{e_2}$, assigned $1$ in a flow instance if and only if there is a pair of occurrences $e'_1$ and $e'_2$ of $e_1$ and $e_2$ such that  $e'_2$ occurs after  $e'_1$ in that flow instance.
\item A binary feature $\ooo{e_1}{e_2}$, assigned $1$ in a flow instance if and only if there is a pair of occurrences $e'_1$ and $e'_2$ of $e_1$ and $e_2$ such that  $e'_2$ occurs before  $e'_1$ in that flow instance.
\end{itemize}

The event occurrence feature $\occ{e}$ can be inferred form the event count feature $\cnt{e}$, since $\cnt{e} = 0$ iff $\occ{e} = 0$, but inclusion of occurrence features can still help ML algorithms to improve root-causing accuracy. In fact, from our experience, occurrence features alone are in most cases enough for root-causing $\pc$ flow failures and we will use them in illustrating examples. 
Recall that we associated events $CLK\_OFF \equiv Clk = 0$ and $CLK\_ON \equiv Clk = 1$ to clock signal $Clk$; now to each of them we associate a binary occurrence feature which we again denote by $CLK\_OFF$ and $CLK\_ON$, respectively. 

The count features are more useful for performance analysis, pre- or post-silicon, and for performance monitoring during in-field operation, say for detecting anomalous activities such as security attacks that have specific performance profiles. There is some similarity in the concept between the count features and profile features used in~\cite{hirsch2020root} in the context of predicting predefined bug types. Since events capture changes in variable values (with some abstraction) and not the duration of the asserted values, the count features capture well signal value switching activities in hardware (which is very important for power analysis), while profiling features are not limited to value switching statistics. On the other hand, the profiling features are limited to only a small fraction of observable variables for which the count features are defined.

There is some redundancy in the in-order features $\ino{e_1}{e_2}$ and out-of-order features $\ooo{e_1}{e_2}$ when they are used in conjunction with the occurrence features. For example,  $\ino{e_1}{e_2}$ can be inferred from $\occ{e_1}$ and $\occ{e_2}$ when $e_1$ or $e_2$ do not occur in the flow instance, or when they occur exactly once in the wrong order, which is also captured by $\ooo{e_1}{e_2}$. 

In general, in-order and out-of-order features can be constructed from more than two flow-event occurrences, say characterizing important sub-flows or stages of the flow, and such features were used in our case studies. In addition, any known correct or wrong order between any sequence of events (not necessarily flow-events) can be added as a feature.

To encode timing information for a subset of events, for each event $e$, and for each pair of distinct flow-events $(e_1, e_2)$ such that $e_1$ comes before $e_2$ in the flow, one can generate:
\begin{itemize}
\item The set of elapsed times $\timeo{e}$ associated to the occurrences of $e$ along the full trace (that is, in all instances of the flow) or along an instance of a flow.  This set of elapsed times, viewed as a distribution, is used to generate distribution features in the same way as for the histograms computed based on counts of events associated to a variable, described earlier. Elapsed times form occurrences of $e$ till the end of the full trace or the trace fragment associated to a flow instance can also be considered.
\item The set of elapsed times $\timet{e_1}{e_2}$ between the occurrences of $e_1$ and $e_2$ in a flow instance or in the entire trace.  This set of elapsed times, viewed as a distribution, is used to generate distribution features in the same way as described above. 
\end{itemize}

Specifically for Power Management flows, the set of expected or actual elapsed times spent in a sleep state or latencies associated to entering or exiting it are among the most relevant timing information useful for root-causing. One can model these directly, or as elapsed times between the events notifying start and end of the modeled time intervals: say the latency of entering a sleep state can be modeled as elapsed time between starting and completing transition to that sleep state. We note that ordering information between pairs of event occurrences $(e_1, e_2)$ can be derived from the elapsed times $\timet{e_1}{e_2}$: negative elapsed times would indicate that $e_1$ occurs after $e_2$ (which can be in accordance with or against the expected order).

Work~\cite{billari2006timing}  considers event ordering features for encoding sequential information in another application domain (a social study), and besides it also encodes timing information, such as the age of individuals at the time of occurrence of
age related events. In our use case, this corresponds to associating to each event the elapsed time from the beginning of the flow, extracted from time stamps attached to event occurrences in the logs. In an experimental evaluation comparing
usefulness of timing, sequential and other information, the authors conclude that the sequential features are the most useful ones for the accuracy of binary classification models in their study. That work does not propose viewing timing information as histograms and analyzing the associated distribution statistics or other properties.
 
We note that when the third requirement from traces, considered in Section~\ref{S.events}, on tracking instance IDs for event occurrences is not satisfied, then sequential and timing features involving two events are more difficult to construct accurately: say when there are multiple sent and received messages, one would require a mechanism to correctly match a sent message with the corresponding received message, otherwise the event ordering and timing features between two event occurrences might not be constructed for the right pairs of event occurrences and root-causing might become less accurate (though as mentioned above, in many cases event occurrence information and also timing information for single events, can compensate for missing event ordering information). Features derived through histograms and distributions for observable variables as described above will still be highly useful.

Useful features are not limited to ones that are associated with variables in the system's software and firmware and with logic signals in the hardware. For root-causing post-silicon test data, the test equipment and environmental effects can also be considered in feature engineering process, in addition to design or manufacturing features associated with integration of a die into packages and boards, and die usage conditions. Sensor data collecting temperature, frequency and voltage information is often very important information for root-causing, and this is true for PM in particular.

\section{The ML Approach}\label{S.ml}

The aim of our method is to explain differences between the passing tests and \emph{subgroups} of failing tests in terms of the information available in the trace logs. The identified subgroups of failures must be such that all failures in a subgroup have the same root cause, and failures in different subgroups represent different root causes. Our method does not require full understanding of all legal behaviors of the system. It  closely follows the intuition behind the debugging steps that the validators follow in root-causing regression failures. We describe how this method works on root-causing PM $\pc$ flow regression test failures.

\subsection{Feature Range Analysis}\label{SS.FRA}

$\RA$ $(\ra$)~\cite{khasidashvili2019range,khasidashvili2021feature} is an algorithm resembling $\RL$ ($\rl$)~\cite{clark1989cn2,furnkranz2012foundations} and $\SD$ ($\sd$)~\cite{klosgen1996multipattern,wrobel1997algorithm,atzmueller2015subgroup}. From the algorithmic perspective, the main distinguishing feature of $\ra$ is that it heavily employs $\FS$ ($\fs$)~\cite{guyon2003introduction} in two basic building blocks of the algorithm: the \emph{ranking} and \emph{basis} procedures. The third basic building block in $\ra$, the procedure called \emph{quality}, is the technique borrowed from $\rl$ and $\sd$, where the selection of rules or subgroups is done solely based on one or more \emph{quality functions}. As a consequence,  $\ra$ can directly benefit from advances in $\sd, \rl$, and $\fs$ algorithms. These three procedures will be explained below.

For a numeric feature $\feat$, $\ra$ defines a \emph{single range feature} as a pair $\rangef \equiv (\feat, value\_range)$, interpreted as a binary feature $\rangef(\sample) \equiv \feat(\sample) \in value\_range$, for any sample $\sample$. And for a categorical feature and level $\level$, the corresponding single range feature is defined as $\rangef(\sample) \equiv \feat(\sample) = \level$. A \emph{range feature} is a single range feature or a conjunction of single range features, also referred to as \emph{range tuples} or simply as \emph{ranges}. A range $\rangef$ \emph{covers} a sample $\sample$ iff  $\rangef(\sample)$ evaluates to true. In the latter case, we will also say that $\sample$ is in $\rangef$ or is captured by $\rangef$. 

The main purpose of $\ra$ is to identify range features that explain positive samples, or subsets (subgroups) of positive samples with the same root cause. The $\ra$ algorithm generates range features that are most relevant for explaining the response, where ‘most relevant’ might mean (a) having a strong correlation or high mutual information with the response, based on one or more correlation measures; (b) explaining part of the variability in the response not explained by the strongest correlating features; or (c) maximizing one or more quality functions, which are usually designed to prefer ranges that separate subgroups of positive samples from negate samples; that is,  a range is preferred if a majority, or perhaps a great majority, of positive samples are covered by the range and a great majority of negative samples fall outside the range. Important examples of quality functions include the ones defined below.

 By treating a range  $\rangef$ as a classifier that predicts the positive class for samples covered by the range and negative class for samples outside the range,  one can apply the well known quality metrics for binary classification to characterize the quality of ranges.  Below we list some of the well known and often used quality metrics for binary classification, where $\TP{\rangef}, \FP{\rangef}, \TN{\rangef}$ and $\FN{\rangef}$ denote \emph{true positive}, \emph{false positive}, \emph{true negative} and \emph{false negative}, respectively, $\AllIn{\rangef}$ denotes the samples count within range $\rangef$, $\Pos$ denotes the positive samples count, and $\All$ denotes the entire samples count. Thus in this context one can interpret  $\TP{\rangef}$ as the count of positive samples covered by the range $\rangef$, $ \FP{\rangef}$ as the count of negative samples covered by the range, $\TN{\rangef}$ as the count of negative samples not covered by the range,  $\FN{\rangef}$ as the count of positive samples not covered by the range, and $\AllIn{\rangef}$ as the count of samples predicted by $\rangef$ as positive. Then the concepts of quality defined for classification tasks can be defined for ranges $\rangef$ as follows:

\begin{itemize}
\item \emph{True Positive Rate} (or\emph{sensitivity}, \emph{recall}, or \emph{hit rate}),   $\TPR{\rangef} = \frac{\TP{\rangef}}{\Pos}$
\item \emph{Predictive Positive Value} (or \emph{precision}) $\PPV{\rangef}  =  \frac{\TP{\rangef}}{\TP{\rangef} + \FP{\rangef}} =   \frac{\TP{\rangef}}{\AllIn{\rangef}}$
\item \emph{Lift} $\Lift{\rangef} = \frac{\TP{\rangef}/\AllIn{\rangef}}{\Pos/\All} =  \frac{\PPV{\rangef}}{\Pos/\All}$
\item \emph{Normalized Positive Likelihood Ratio}  $\NPLR{\rangef} =  \frac{\TPR{\rangef}}{\TPR{\rangef}+ \FPR{\rangef}}$,  where $\FPR{\rangef} = \frac{\FP{\rangef}}{\FP{\rangef} + \TN{\rangef}}$
\item \emph{Weighted Relative Accuracy} $\WRAcc{\rangef}= \frac{\AllIn{\rangef}}{\All}\cdot (\frac{\TP{\rangef}}{\AllIn{\rangef}}-\frac{\Pos}{\All}) =  \frac{\AllIn{\rangef}}{\All}\cdot (\PPV{\rangef}-\frac{\Pos}{\All}) $ 
\item \emph{ROC Accuracy} associated with \emph{Receiver Operating Characteristic} (ROC) \\  $\ROC{\rangef} = \TPR{\rangef}-\FPR{\rangef}$
\item \emph{$F_1$-score} $\FOS{\rangef} = \frac{2\cdot \PPV{\rangef}\cdot\TPR{\rangef}}{\PPV{\rangef}+\TPR{\rangef}}$
\item \emph{Accuracy} $\Acc{\rangef} = \frac{\TP{\rangef} + \TN{\rangef}}{N}$
\end{itemize}

The RA algorithm works as follows:

\begin{enumerate}
\item The $\ra$ algorithm first ranks features highly correlated to the response; this can be done using an ensemble \emph{Feature Selection}~\cite{guyon2003introduction} procedure, we refer to it as \emph{ranking} procedure and denote the number of features it selects by $\rpafeat$ (this number is controlled by user). In addition, $\ra$ uses the $\mrmralgo$~\cite{de2013mrmre} algorithm to select a subset of features which both strongly correlate to the response and provide a good coverage of the entire variability in the response; this algorithm selects a subset of features according to the principle of \emph{Maximum Relevance and Minimum Redundancy (\MRMR)~\cite{ding2005minimum}}, we refer to this procedure as \emph{basis} and denote the number of features it selects by $\rpamrmr$ (this number too is controlled by user). 
\item For non-binary categorical features selected using the ranking and basis procedures described above, or optionally, for all non-binary categorical features, for each level a binary range feature is generated through the one-hot encoding. In a similar way, a fresh binary feature is generated for each selected numeric feature and each range constructed through a discretization procedure. These features are called single-range features. Unlike $\rl$ and $\sd$, in $\ra$ the candidate ranges can be overlapping, which helps to significantly improve the quality of selected ranges. The $\ra$ algorithm then applies ranking and basis procedures to select $\rpafeat$ and $\rpamrmr$ most relevant single ranges, respectively; in addition, $\ra$ selects single range features that maximize one or more quality functions -- $\rpafeat$ single rage features per quality function used. We refer to the quality-function based selection of ranges as \emph{quality}.
\item For each pair (or optionally, for a subset of pairs) of selected single range features associated with different original features, $\ra$ generates range-pair features which have value $1$ on each sample where both the component single-range features have value $1$ and have value $0$ on the remaining samples. $\ra$ then applies ranking, basis and quality procedures to select respectively $\rpafeat$ and $\rpamrmr$ most relevant range pairs. 
\item Similarly, from the selected single ranges and selected range pairs, the $\ra$ algorithm builds range triplets, and applies ranking, basis and quality procedures to select most relevant ones. Ranges with higher dimensions can be generated and selected in a similar way. 
\end{enumerate}

$\ra$ is implemented in Intel's auto-ML tool EVA~\cite{khasidashvili2019range,khasidashvili2021feature}. It has been used for root-causing and design space exploration also in other areas of microprocessor design~\cite{manukovsky2020machine,manukovsky2021machine}.

\subsection{Quality orders}\label{SS.FRA.ord.}

For our root-causing application, the first priority is to find range tuples with high precision -- preferably with the maximal precision $1$, since such a range tuple provides a clear explanation for the subset of positive samples covered by it (this explanation might or might not be the 'true' explanation). The second priority is to find high precision ranges with high sensitivity (that is, high coverage of positive samples), since that way one can identify root-causing hints valid for many failures thus these failures might be solvable with the same fix -- as a subgroup of failures with the same root cause. In addition, in general, range tuples with smaller dimension might be preferable among ranges with the same precision and coverage, as they give simpler explanation for the covered positive samples. Therefore, given a dataset with a binary label, that is, given $\All$ and $\Pos$, it is convenient to view that the precision $\PPV{\rangef}$ and coverage $\AllIn{\rangef}$ uniquely determine $ \TP{\rangef},  \FP{\rangef}, \TN{\rangef}$ and $\FN{\rangef}$, thus all the quality functions above are expressible through precision and coverage.

In this work, in addition to quality function based selection, we suggest to use \emph{quality orders} for prioritizing ranges for selection. In particular, we define \emph{Coverage Weighted Precision (CWP)}, or \emph{Weighted Precision} for short, as a lexicographic order on pairs  $(precision, coverage)$. That is, $\CWP{\rangef_1} > \CWP{\rangef_2}$ if and only if $\PPV{\rangef_1} > \PPV{\rangef_2}$ or $\PPV{\rangef_1} = \PPV{\rangef_2} \wedge \AllIn{\rangef_1} >   \AllIn{\rangef_2} $. For binary responses, which is the focus of this work, \emph{CWP} induces the same ordering as  \emph{Positive Coverage Weighted Precision (PCWP)} defined  as a lexicographic order on pairs  $(precision, positive\_coverage)$, where $positive\_coverage$ is the count of positive samples covered by the range feature. For numeric responses, there might be multiple suitable ways to define positive samples (see Section~6.2 in~\cite{khasidashvili2021feature}), thus there will be multiple useful ways to define \emph{PCWP} orderings. While $WRAcc$ prefers selection of ranges with high precision and high coverage, the ordering induced by $WRAcc$ is not equivalent to weighted precision. For example, consider dataset with $\All=80$ samples containing $\Pos=20$ positive ones, and consider two range tuples $\rangef_1$ and $\rangef_2$,  covering $\TP{\rangef_1}=10$ positive and $\FP{\rangef_1}=0$ negative samples, and $\TP{\rangef_2}=13$ positive and $\FP{\rangef_2}=7$ negative samples, respectively. Then $\PPV{\rangef_1} = 1$ is significantly higher than $\PPV{\rangef_2} = 0.65$, thus  $\CWP{\rangef_1} > \CWP{\rangef_2}$, while  $\WRAcc{\rangef_1} = 0.09375$ is smaller than $\WRAcc{\rangef_2} = 0.1$.  

\emph{Coverage Weighted Lift} and  \emph{Positive Coverage Weighted Lift} are defined similarly, and they define the same orderings as \emph{CWP} and \emph{PCWP}, respectively. Among  two range tuples with the same $WRAcc$, one with smaller coverage has higher precision, therefore \emph{Coverage weighted WRAcc} is defined as the lexicographic order on pairs $(WRAcc, coverage)$, while \emph{Precision weighted WRAcc} is defined as as the lexicographic order on pairs $(WRAcc, 1/coverage)$, or equivalently, on pairs $(WRAcc, precision)$. One can refine the remaining quality functions in the same way. The dimension of the range tuples can be added to these lexicographic orders as the third component in order to ensure that range tuples with smaller dimensions are selected first among tuples with the same precision and coverage.

\subsection{Basic Root-Causing Step}\label{SS.basic}

We explain the main idea of how $\ra$ helps root-causing on an example of Figure~\ref{register_x_167}, which displays a binary feature $CLK\_OFF$, associated to event $Clk = 0$, ranked highly by $\ra$ and plotted for validator's inspection. The dots in the plot, known as a \emph{Violin plot}, correspond to rows in the input dataset, thus each dot represents a passing or failing instance of $\pc$ flow. Value $1$ on the $Y$ axis (the upper half of the plot) represents the failing tests and value $0$ (the lower half) represents the passing ones.
The plot shows that when $CLK\_OFF = 0$, that is, when signal $Clk$ was never set to $0$ along the execution of the flow instance, all tests fail! This is a valuable information that the validator can use as an atomic step for root-causing.  The legend at the top of the plot specifies that there are actually $115$ failing tests and $0$ passing tests with $CLK\_OFF=0$, and there are $85$ failing tests and $247$ passing ones with $CLK\_OFF=1$. 

Let's note that the range feature $(CLK\_OFF, 0)$ has precision $1$, as defined in Subsection~\ref{SS.FRA}, since it only covers positive samples (which correspond to failing tests). To encourage selection of range features with precision $1$ in root-causing runs,  in $\ra$ we use weighted precision along with \emph{WRAcc} and \emph{Lift}, which also encourage selection of range features with precision~$1$.

\subsection{Root-Causing Strategies}\label{SS.strategies}

During different stages of product development, the nature of validation activities might vary significantly. When there are many failing tests, the root cause for failures might be different for subgroups of all failed tests. In this case it is important to be able to discover these subgroups of failing tests so that failures in each subgroup can be debugged together to save validation effort. On the other hand, it is important to be able to efficiently debug individual failures. In our approach, the strategies of identifying subgroups of failing tests with similar root cause and putting focus on root-causing individual failures correspond to breadth-first and depth-first search strategies in the \emph{Root-Causing Tree ($\rct$)} that we define next.

\begin{figure}[t]
\center
\begin{subfigure}{4.4cm}
\includegraphics[width=\textwidth=0.9]{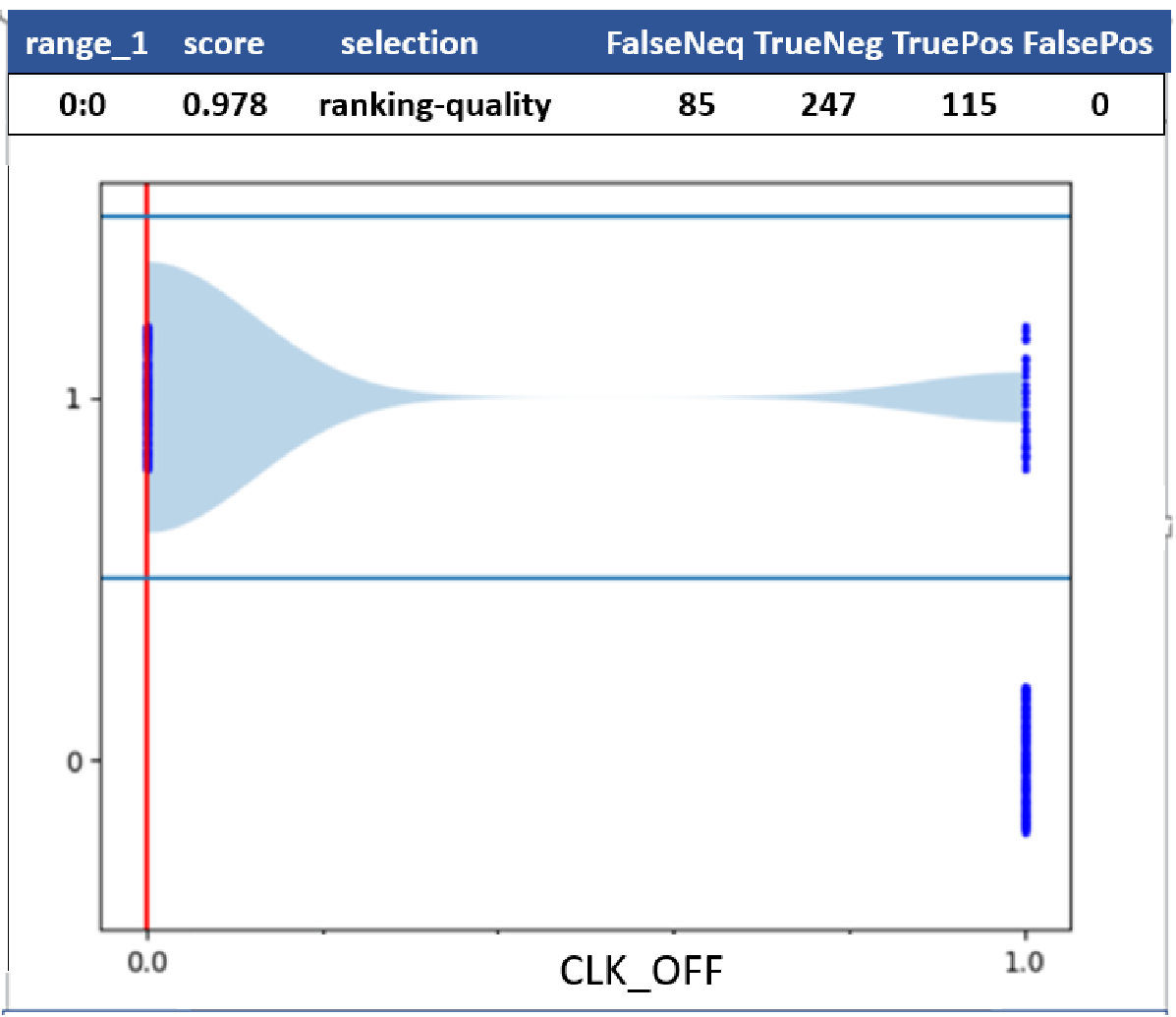}
\caption{Range feature ranked highly by $\ra$.} 
\label{register_x_167}
\end{subfigure}
\rule{0px}{100px}
\begin{subfigure}{7.2cm}
\includegraphics[width=\textwidth=0.9]{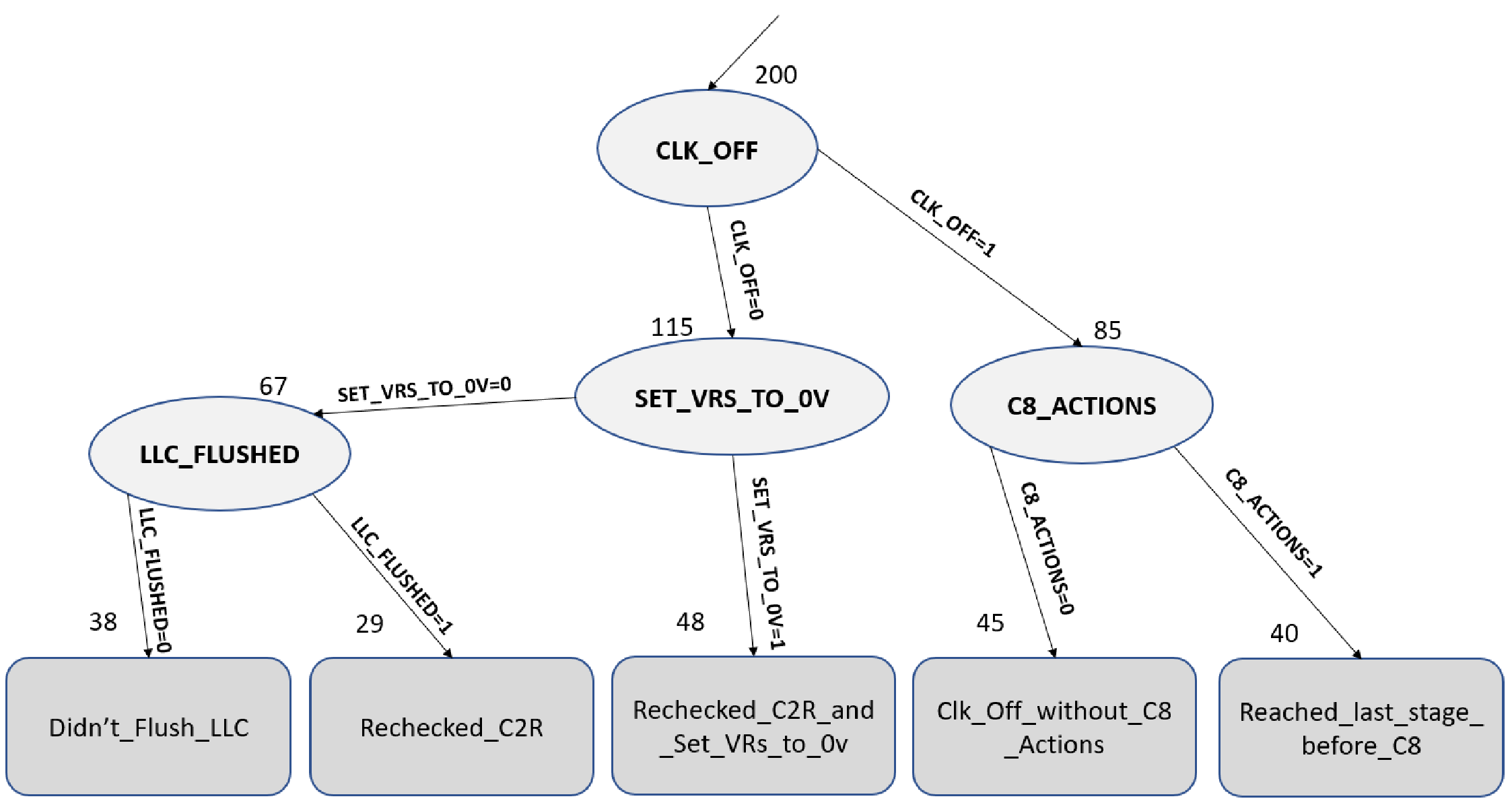}
\caption{Root-causing tree of failures of PkgC8 and wakeup-timer.} 
\label{wakeup_timer_buckets}
\end{subfigure}
\caption{Iterative root-causing procedure.} 
\label{rct_procedure}
\end{figure}

Our proposed procedure for root-causing regression failures is an iterative process and requires a guidance from the validator -- or inputs from an \emph{oracle}, to make the formulation of $\rct$ procedure formal -- in order to decide whether the ranked root-causing hints are \emph{sufficient to conclude the root cause} or further root-causing effort is required, including \emph{upgrading the initial dataset} with more features according to what has been learned from the root-causing procedure that far. Generation of the ranked list of root-causing hints in each iteration of building $\rct$ (splitting a node in the current tree) is fully automated by using the $\ra$ algorithm, but unlike the $\rl$ and $\sd$ algorithms, the input data to the $\rct$ procedure can change from one iteration of $\ra$ algorithm to a next iteration. As a result, in $\rct$ procedure user guidance is explicit while data preparation and inspection of results are outside of the scope of $\rl$ and $\sd$ algorithms (thus  $\rl$ and $\sd$ are deemed as fully automated even if the results are not satisfactory and rerun is required with a more appropriate dataset and/or different parameter values of $\rl$ and $\sd$ algorithms). 

The iterative root-causing procedure is illustrated using Figure~\ref{wakeup_timer_buckets}. More specifically, Figure~\ref{wakeup_timer_buckets} presents an $\rct$ of regression failures of $\pc$ with random wakeup-timer, which will issue  an interrupt to trigger a wake-up out of $\pc$ flow. Using the procedure described next, after a few iterations, at the leaves of the tree we get subgroups of failures characterized by the length of the wakeup-timer and which actions were achieved. Initially, in the regression we see $200$ failing tests as one group of failing tests, and there are $247$ passing tests. 

\begin{itemize}
\item[(a)] [\emph{Initialization}] 
Initialize the root-causing tree $\rct$ to a node $n$, mark it as open, set the input dataset $D$ passed to the iterative root-causing procedure to the dataset engineered from traces as described in Section~\ref{S.features}, and set the number of iterations $i = 0$. Go to step~$(c)$.
\item[(b)] [\emph{Iterations}]
In this step, the procedure receives a four-tuple $(\rct, n, S, D_S)$ where $\rct$ is the current tree, $n$ is a node in it selected for splitting, $S$ is the subgroup of failing tests associated to node $n$, and $D_S$ is the dataset to be used for splitting the subgroup $S$, with the aim to arrive to smaller high quality subgroups with different root cause or root causes each.  
 $\ra$ is performed on $D_S$ and the oracle chooses the best range feature to split the tree at node $n$. (Any range feature not identical to the response implies a split at $n$.) We add two sons to $\rct$: one will be associated with the subgroup of failing tests covered by the selected range feature, and the other one with the subgroup containing the rest of failing tests from $S$; both sons are marked as open. Increment $i$ and go to step $(c)$.
\item[(c)] [\emph{Decisions}]
\begin{enumerate}
\item If all leaves of $\rct$ are marked as closed, exit. Otherwise go to $(c).2$:
\item Choose a leaf node $n$ not marked as closed -- this is the leaf targeted for splitting in order to refine the associated subgroup $S$ of failing tests. if the oracle decides there is no need to split $S$, that is, all failing tests in $S$ can be associated to a single root cause or $S$ is empty, then mark the leaf $n$ as closed and go to $(d)$. Otherwise go to $(c).3$:
\item Define $D_S$ as follows: if $i > 1$, from the input dataset $D$ subset rows that correspond to failing tests in this subgroup and rows corresponding to the passing tests, to obtain dataset $D_S$. Drop from $D_S$ all features that were used for splitting at the parent nodes; in addition, drop the features that are identical to the dropped features, as well as features that are identical to the response (the class label), in $D_S$. Otherwise (case $i = 0$), set $D_S = D$. Go to $(c).4$.
\item  If there are no features in $D_S$ and no new features can be added, mark the leaf node $n$ as closed and go to step $(c)$. Otherwise, if $D_S$ contains all features relevant to $S$ (subject to the oracle's decision), or no additional features are available, go to step $(b)$ with four-tuple $(\rct, n, S, D_S)$. Otherwise go to step $(e)$ with four-tuple $(\rct, n, S, D_S)$. 
\end{enumerate}
\item[(d)] [\emph{Knowledge Mining}] This step performs mining of knowledge learned along a closed branch of the root-causing tree, and it will be discussed further in Subsection~\ref{SS.mining}. Then go to step~$(c)$.
\item[(e)] [\emph{Data Refinement}] This step corresponds to the situation when the oracle wants to add more features that can potentially help to split subgroup $S$ further, and it will be discussed further in Subsection~\ref{S.refinement}.  Then go to step~$(b)$ (with an updated dataset $D_S$).
\end{itemize}

Let's note that in the root-causing tree of Figure~\ref{wakeup_timer_buckets}, event $C8\_ACTIONS$ is not a flow-event -- it does not occur in $\pc$ flow description. It helps to explain why some of the flow instances that successfully go through the $CLK\_OFF$ stage of the $\pc$ flow fail to complete the flow. Also, the events of the $\pc$ that should occur between $CLK\_OFF$ and  $CORE\_WAKE\_TO\_C0$ do not occur in the root-causing tree. The root-causing tree therefore provides pretty neat and succinct explanations and grouping of the failures, and gives deeper insights compered to  just identifying where in the flow the flow instance starts to deviate from the expected execution. As discussed in Section~\ref{S.features}, the information that event $C8\_ACTIONS$ should occur before events notifying exit stages of $\pc$ flow can be mined as out-of-order features, optionally. During building the tree in Figure~\ref{wakeup_timer_buckets}, this information was not coded as a feature in the input dataset~$D$.

In each splitting iteration, the range feature selected for splitting can be multi-dimensional, such as pairs or triplets of feature-range pairs. That is, in general splitting is performed based on a set of events, not necessarily based on just one event. This will be discussed in more detail in Subsection~\ref{SS.tuples}. Each iteration is zooming into a lower-level detail of the flow execution and smaller set of modules of the system. 

We note that the features identical to the response in dataset $D_S$ that are dropped from analysis in step $(c)$ are very interesting features for root-causing as they fully separate the current subgroup $S$ and the passing tests.

Our iterative root-causing procedure assumes that there are passing tests. This is a reasonable assumption but does not need to hold at the beginning of a design project. In such cases, unsupervised rule based or clustering algorithms can be used.

Finally, let's note that at every iteration in the proposed root-causing procedure, we select only one range feature to perform a split. However, multiple range features can be selected and splitting can be performed in parallel (and then the root-causing tree will not be binary) or in any order (then the root-causing tree will stay binary and in each iteration we add more than one generation of children to the tree). This reduces the number of required iterations.

\subsection{Root-Causing Using Range Tuples}\label{SS.tuples}

\begin{figure}[t]
\begin{subfigure}{3.8cm}
\includegraphics[width=\textwidth=0.9]{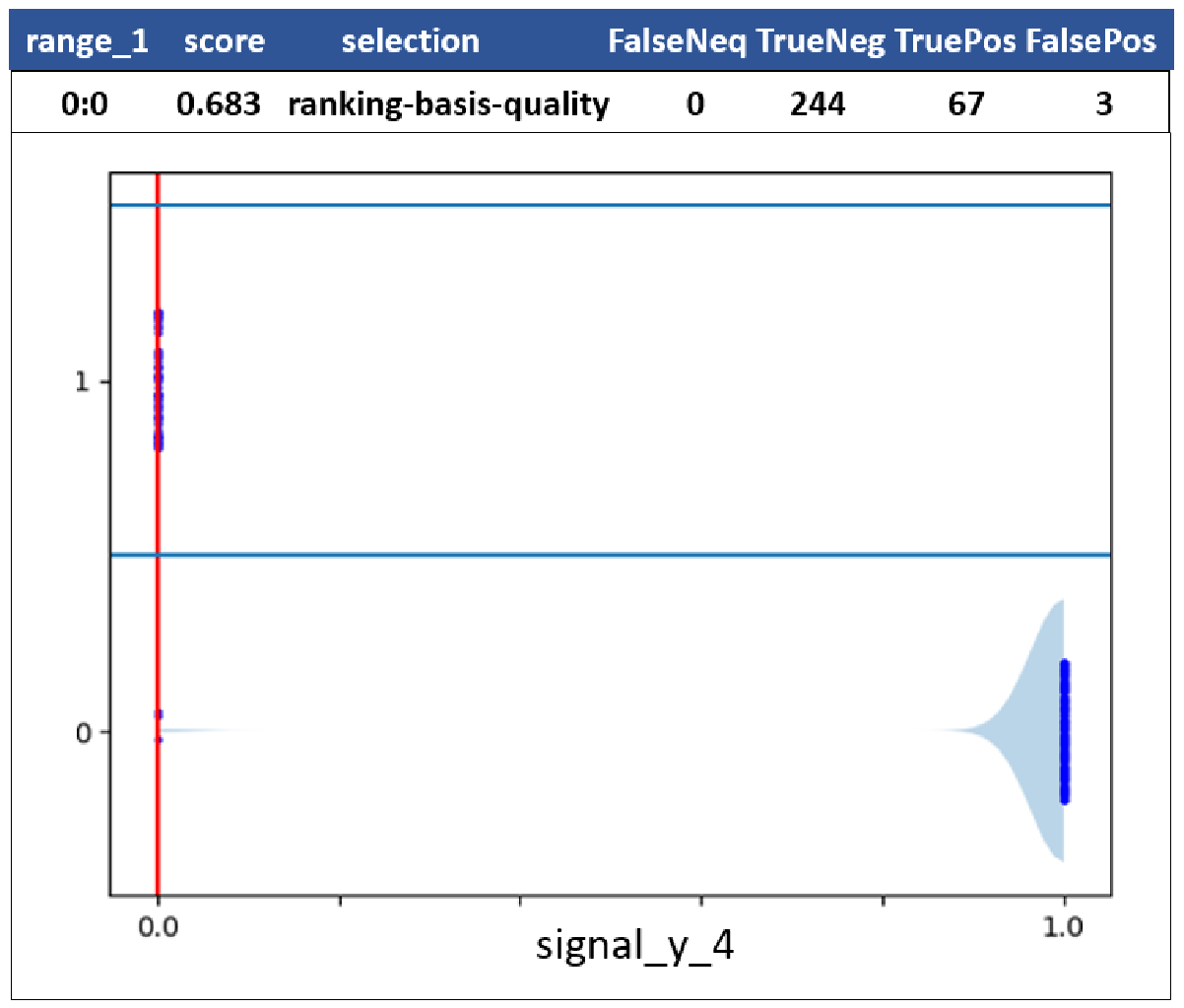}
\caption{Highly ranked single range feature $signal\_y$  cannot isolate failing tests from the passing ones.} 
\label{signal_y_4}
\end{subfigure}
\rule{0px}{100px}
\begin{subfigure}{3.7cm}
\includegraphics[width=\textwidth=0.9]{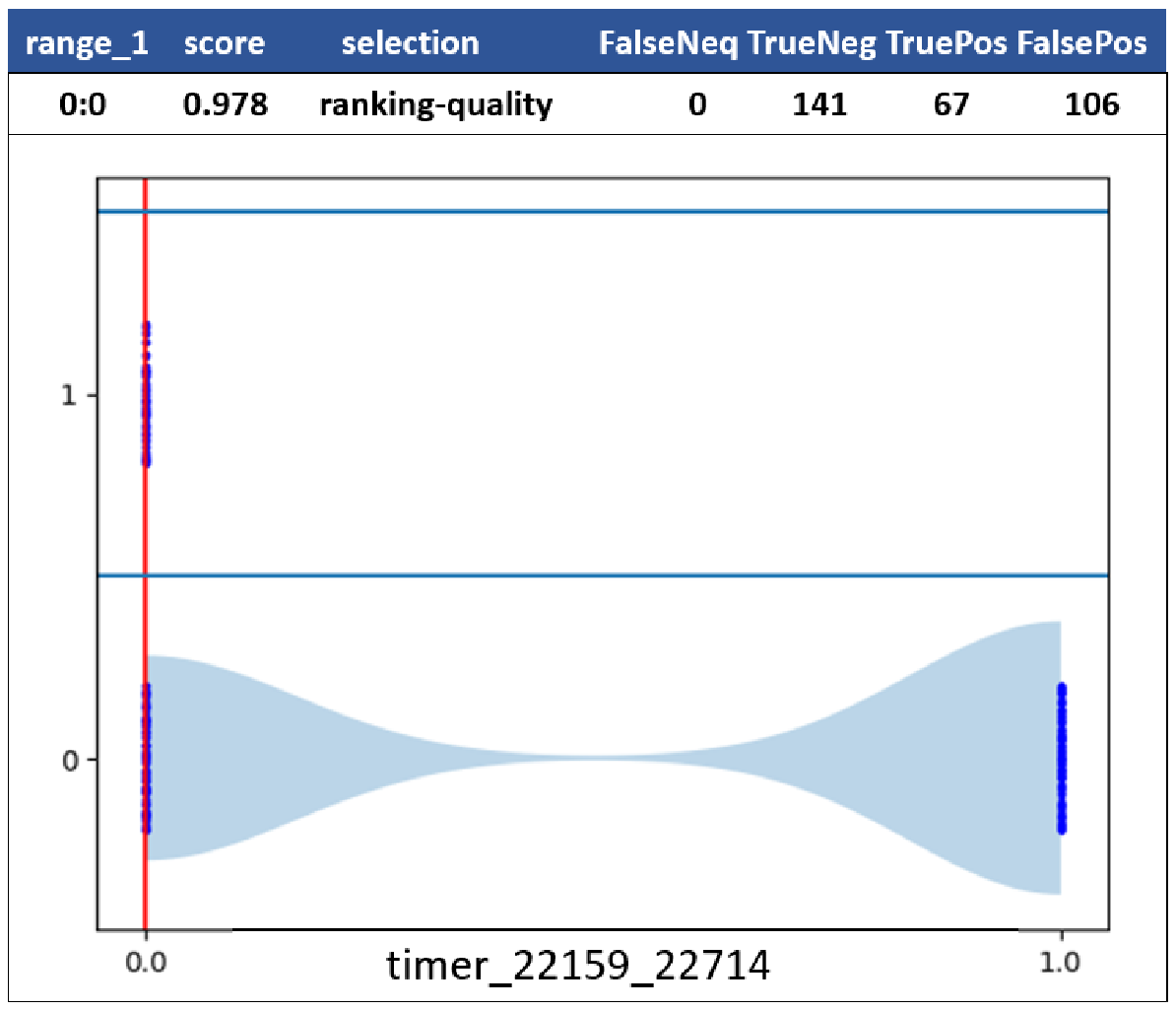}
\caption{Single range  feature $timer\_22159\_22714$ alone cannot isolate failing tests from the passing ones.} 
\label{timer_range}
\end{subfigure}
\rule{0px}{100px}
\begin{subfigure}{3.9cm}
\includegraphics[width=\textwidth=0.9]{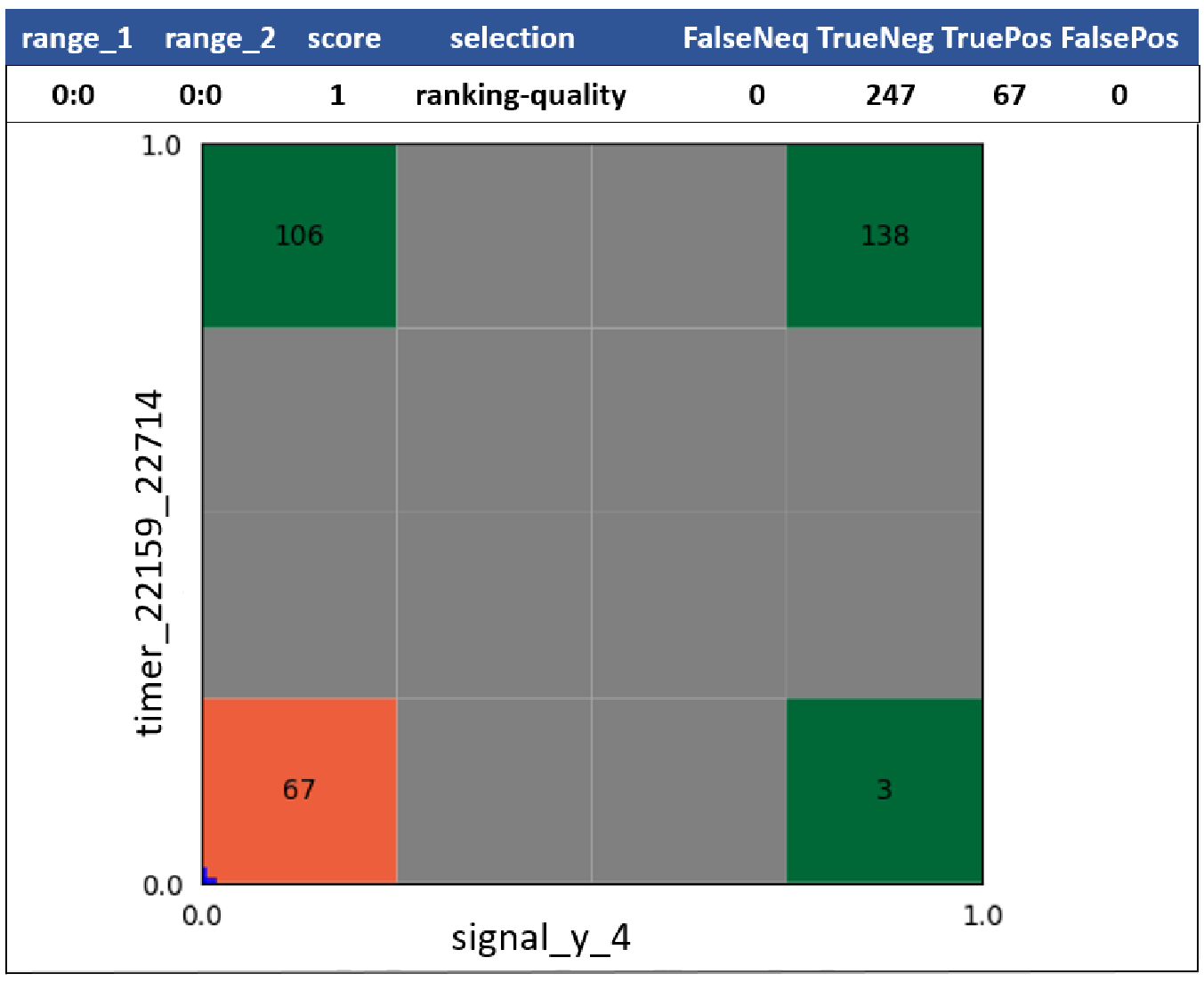}
\caption{Range pair that combines single ranges (a) and (b) isolates failing tests from the passing ones.}  
\label{range_pair}
\end{subfigure}
\caption{Usefulness of range tuple features in iterative root-causing procedure.}
\label{range_tuples}
\end{figure}

Consider the feature $signal\_y\_4$ in Figure~\ref{signal_y_4}, associated to event $signal\_y = 4$, and feature $timer\_22159\_22714$ in Figure~\ref{timer_range}, associated to event $timer \in [22159, 22714]$. While for a big majority of tests ($67$ out of $70$, in this case) the value $0$ of feature $signal\_y\_4$ implies test failure, there are a few tests ($3$ out of $70$, in this case) that are passing. So, the single range feature  $\rangef_1 = (signal\_y\_4, 0)$ cannot fully isolate the failing tests from passing ones. As can be seen from Figure~\ref{timer_range}, the second single range feature $\rangef_2 = (timer\_22159\_22714, 0)$  cannot isolate failing tests from the passing tests either – along with $67$ failing tests, there are $106$  passing tests when timer does not occur at time range $[22159, 22714]$. However, the range pair feature $\pairf = \rangef_1 \wedge \rangef_2$ (the conjunction of two binary features $\rangef_1$ and $\rangef_2$, as defined in Subsection~\ref{SS.FRA}) fully isolates the $67$ failing tests from the passing tests, because the $3$ negative samples covered by $\rangef_1$ and $106$ negative sample covered by $\rangef_2$ are disjoint. The range pair feature $\pairf$ is depicted in Figure~\ref{range_pair}, where the color encodes the ratio of positive and negative samples within the colored squares: the red color encodes positive samples and green encodes negative samples. 

Usage of range pairs and range tuples of high dimensions can accelerate debugging as they allow to arrive to a root cause in fewer debugging iterations. On the other hand, the splittings of subgroups of failures associated to high dimensional range tuples might not match the intuition of the validator (the oracle's decision) controlling the iterations in the root-causing tree. Thus the time spent by $\ra$ algorithm (or other $\rl$ or $\sd$ algorithms which can be used instead of $\ra$ at each splitting iteration) will be wasted. One useful heuristic of choosing ranges for next splitting iterations is to choose the range features with precision $1$ that originate from the flow-events as a default, without intervention of the validator, if such range features are selected by $\ra$ as candidates for splitting. 

To aid the validator in choosing the range features that are associated to well understood events, to every ranked candidate range features we associate a list of very highly correlated or identical features, with respect to $D_S$. Thus, instead of a highly ranked feature the validator can chose another representative with a clear interpretation.  These strongly correlated or identical features are computed by $\ra$ anyway, as a means to reduce the redundancy in the identified subgroups and to speed-up the algorithm~\cite{khasidashvili2019range,khasidashvili2021feature}. In $\ra$, the usage of the $\MRMR$ procedure helps to not miss interesting range features for splitting, when there are many features in the analysis. Finding diverse, non-redundant and concise sets of subgroups is an important direction in the $\sd$ research~\cite{lavrac2004subgroup,van2012diverse}, and is very relevant in the context of minimizing the validator's involvement as the decision maker in controlling the root-causing iterations.

\subsection{Root-Causing in Hierarchical Order}\label{S.refinement}

For large systems with many modules, it might not be wise or feasible to generate upfront all features that can be engineered from traces, and include them into the initial dataset $D$ used in the first root-causing iteration. The root-causing iterations can also be performed in a hierarchical manner, and this is actually a preferred option which we adopted in our experiments. 

In the hierarchical approach, during the initial iterations, one can choose to only consider features associated with some of the important modules as well as with observable signals on module interfaces, in order to include in analysis the features associated to all the flow-events in the flow of interest. And if root-causing hints collected so far indicate that the bug could be in a particular module or modules, features originating from these modules can be added to analysis (and optionally, features from some other modules can be dropped).  This flexibility of refining the input data during root-causing iterations helps to keep the datasets analyzed by ML algorithms smaller and focused, and speeds up the root-causing iterations. In particular, when the iterative procedure arrives at step $(e)$, one could consider adding  features originating from relevant modules if they were not present, or consider engineering more features associated to additional variables from the relevant modules, when possible, and then continue with step $(b)$, with tuple $(\rct, n, S, D_S)$ received at step (e), but with the updated $D_S$. As an example, in the root-causing tree of Figure~\ref{wakeup_timer_buckets}, the rightmost branch was closed with label $Reached\_last\_stage\_before\_C8$, because data from the respective module was not available at the time of initial experiments.  Interestingly, when the relevant data became available and was added to analysis, further root-causing revealed that the corresponding regression test failures were not caused by a bug, instead, the self-checks in these tests were incorrect, and were fixed as a result of the gained insights during the root-causing iterations. 

We note that there might be other useful ways to revise a current dataset, not covered by our algorithm, and it is not our goal to be exhaustive in that respect. For example, as mentioned in Section~\ref{S.features}, feature engineering depends on the definition of the trace fragment associated to a flow instance, and our choice in this work was to define the relevant trace fragment upfront as part of data preprocessing step prior to $\rct$ construction, while we could make the definition of relevant trace fragment as part of the $\rct$  procedure, since identification of the most relevant fragments of trace to engineer features from it is also affected by the learning acquired as part of root-causing iteration performed so far.

\subsection{Mining Rules}\label{SS.mining}

In order to make root-causing easier for validators and to enable leveraging the knowledge accumulated during root-causing activities, we propose to mine validation knowledge as follows:
\begin{itemize}
\item Maintain and improve a table associating their meaning with well understood events.  This table is used to display the meaning of the events and event combinations associated to range features ranked highly by $\ra$ algorithm when the latter are presented to the validator to choose the best root-causing hints and decide on next root-causing iteration.
\item Construct and improve a database or rules as follows: At step $(d)$ of the root-causing procedure, before returning to step $(c)$, the validator can assign a label to the respective subgroup of failures, and add a rule to the database of rules associated to the branch of the tree leading to that subgroup.  The antecedent of this rule is the conjunction of conditions associated with the ranges along this branch and the consequent is the label of the subgroup $S$ at the leaf of this branch. As an example, the rule associated to the branch ending at node $Clk\_Off\_without\_C8\_Actions$ is written as: $$CLK\_OFF = 1 \wedge C8\_ACTIONS = 0 \longrightarrow Clk\_Off\_without\_C8\_Actions$$
\end{itemize}

The learned rules in the rule database can be used to predict root cause of test failures in future regressions by building predictive models from rules, using the algorithms developed under $\rl$ or related algorithms such as~\cite{lavrac2004subgroup} that combine $\rl$ and $\sd$ techniques for building predictive models from rules.  An alternative approach is to train a model predicting the learned failure labels (failure labels learned during $\rct$ construction) once a rich set of failure labels and dataset $D$ engineered from passing tests and failing tests with these failure types are available, like this is done in previous work~\cite{mammo2016bugmd,hirsch2020root} for predefined failure labels. In addition to the features in $D$, the range features corresponding to the left-hand sides of the mined learned rules can be used as derived features when training the model, since this can improve the quality of the trained model as was demonstrated in~\cite{khasidashvili2021feature}.

It is important to point out that the learned failure labels and rules added to the database are based on the intended behavior of the system and not on its current implementation. Thus, when the design is updated (say due to bug fixes), the design intent usually stays the same. Therefore, the failure labels and rules stay relevant as the design evolves, as long as the design intent is not changed. That is, the rules are more like the reference models or other types of assertions, but have the opposite meaning in that they capture cases of incorrect behavior rather than the intended behavior.

\subsection{Backtracking and Termination of Root-Causing Procedure}

In our root-causing procedure, we did not discuss backtracking during the $\rct$ construction. When backtracking is not used, each subgroup of failing tests is split at most once, which can be done in a finite number of ways (in binary $\rct$ construction procedure, a subgroup is split in two subgroups or not split at all). Furthermore, steps $(a), (b), (d)$ transition to step $(c)$ which starts with a termination check, and step $(e)$ transitions to $(b)$ which then transitions to $(c)$. Therefore, assuming backtracking is not used and assuming step (b) succeeds by selecting a range that is used for splitting, it is straightforward to conclude that the $\rct$ construction will terminate (there will be no infinite loops). 

In general, it can be the case that step (b) does not generate valuable root-causing hints and the validator cannot refine the subgroup at hand further. The most likely reason for this is that the used dataset does not contain relevant features. If that is the case, it is useful to backtrack to the same subgroup $S$ but upgrade the dataset $D_S$ by adding potentially relevant features, by going through step $(e)$. Otherwise, the problem might be due to incorrect splitting decisions taken earlier during $\rct$ constriction or poor quality of top ranked range features returned by $\ra$, and in such cases too backtracking to an already visited subgroup might be useful. Assuming the number of features available for root-causing is limited and only a limited number of backtracking to an already visited subgroup is allowed, $\rct$ construction will always terminate.

\section{Results}\label{S.results}

We explored the $\RA$ algorithm for root-causing Power Management protocol failures on several regressions, starting with a synthetic regression of $\pc$ and random wakeup-timer (Figure~\ref{wakeup_timer_buckets}), and continuing with authentic random PM regressions on a FullChip emulation platform, on two recent Intel products. We found this method very effective for classifying the failures and providing valuable root-causing hints. It also significantly improves the productivity of the validators as the most difficult and time consuming root-causing activities are fully automated. 

In Table~\ref{bench_stats} we report statistics of datasets used in the experiments. The first column lists eight datasets engineered from $PM$ test regression logs and fed to $\ra$ for generating root-causing hints. The next two columns $fail$ and $pass$ report counts of failing and passing tests in a regression run, respectively. The column $features$ reports the number of features engineered from trace logs, and the column $vars$ reports the number of observable variables (signals from hardware, variables in the software or firmware, parameter values collected from sensors, or other parameters) to which the features are associated with. The column $count$  reports the number of count features and variables to which they are associated with, separated with ``/'', and similarly the columns $occurs,  timing$ report the number of occurrence and timing features and the variables to which they are associated with, separated with ``/'', respectively. Finally, the column $ordering$ reports the count of features associated to in-order and out-of-order sequences of two or more events; and the column $other$ reports the count of other features relevant for root-causing (e.g., features associated with parameters that do not change along test but might change from test to test, such as particular IDs, modes and conditions of operation). Note that we associate histograms with count and timing features and their count is the same as the number of variables reported in columns $count$ and  $ timing$; this table does not cover counts of features associated with these histograms (such as histogram statistics $min$, $max$, $mean$, ...), used in root cause analysis.

\begin{table*}
\begin{tabular}{rrrrrrrrrr}
\toprule
 data &  fail &  pass &  features &  vars &        count &       occurs  &      timing  &  ordering  &  other \\
\midrule
Data1 &   237 &   247 &       932 &   472 &     368/158 &      514/302 &        50/12 &            0 &         0 \\
Data2 &   237 &   246 &       932 &   472 &     275/107 &      607/353 &        50/12 &            0 &         0 \\
Data3 &   237 &   247 &       492 &   156 &         0/0 &      424/148 &         36/8 &           26 &         6 \\
Data4 &   226 &    44 &       867 &   178 &         0/0 &      793/170 &         40/8 &           27 &         7 \\
Data5 &   463 &   291 &       869 &   178 &         0/0 &      797/170 &         38/8 &           27 &         7 \\
Data6 &  2373 &   142 &       908 &   206 &         0/0 &      835/198 &         38/8 &           27 &         8 \\
Data7 &  2626 &   200 &       927 &   207 &         0/0 &      854/199 &         38/8 &           27 &         8 \\
Data8 &  2626 &   200 &       944 &   207 &         0/0 &      870/199 &         39/8 &           27 &         8 \\
\bottomrule
\end{tabular}
\caption {Benchmark dataset statistics}
\label{bench_stats}
\end{table*}

The hierarchical approach to root-causing allows to keep the datasets fed to $\ra$ pretty small. While $\ra$ can handle input datasets with thousands of columns and millions of rows, it is important to keep the datasets small to speed up the iterations. As it can be seen from Table~\ref{bench_stats}, in our experiments the number of features range from a few hundred to a thousand.
The number of rows range from around few hundreds to a few thousands. The largest analyzed dataset $Data8$ has $944$ features and $2826$ rows. On such small datasets, each iteration would typically take up to $2$ minutes. Number of iterations range from just one to ten or so, with the depth of the $\rct$ reaching up to four or five levels on average.

To the best of our knowledge, this work represents the first usage of $\sd$ and $\rl$ techniques towards automating root cause analysis in validation domain. Therefore, in the experimental  evaluation below on PM datasets, our first goal is to demonstrate that the root-causing hints identified by our approach are of high quality according to widely accepted objective measures of quality (Tables~\ref{largest_dataset_range_ranking} and \ref{pos_coverage_ppv_all_one}). In addition, we take a closer look at some of the $\sd$  heuristics that help to diversify the generated root-causing hints and evaluate them on PM datasets (Tables~\ref{pos_coverage_ppv_all_one} and~\ref{pos_coverage_gamma}). Methods used for feature engineering from traces is of critical importance for the entire approach to succeed, and therefore we also analyze which types of features engineered from trace logs are most useful in generating root-causing hints (Table~\ref{hist_features_ratio}).  The core $\ra$ algorithm described in Section~\ref{SS.FRA} is not a contribution of this work and while we mostly use $\ra$ implementation in Intel's EVA tool in our experiments, it is not our aim to demonstrate that it performs better than $\sd$ implementations on PM datasets (recall that $\ra$ leverages $\sd$ heuristics in one of its main building blocks). In Tables~\ref{pos_coverage_ppv_all_one} and~\ref{pos_coverage_gamma}, we include results obtained using the pysubgroup package for $\sd$~\cite{lemmerich2018pysubgroup}, to give an idea about the relative performance, on PM datasets, of $\ra$ compared to a basic implementation of $\sd$. (The pysubgroup package is not restricted to non-commercial usage and is easy to use;  it does not support many advanced heuristics of $\sd$.) A more comprehensive  experimental evaluation of $\ra$ with $\sd$ and $\rl$ on Intel as well as public datasets can be found in~\cite{khasidashvili2019range,khasidashvili2021feature}.

\subsection{Quality Metrics on the Largest PM Dataset}

Table~\ref{largest_dataset_range_ranking} reports $\ra$ results in the original root-causing activity after the first $\rct$ iteration on the dataset $Data8$ with $944$ features and $2826$ rows.  Each row in the table corresponds to a selected range feature (the names of features and the selected ranges are not shown), and the first column $Dim$ shows the dimension of the range features -- range features up to triplets were generated. 
Two heuristics for generating diverse set of ranked range features were used: (1) the number of ranges selected per feature was limited to $2$ (this is a default value and was used in all experiments); and (2) among the identical range features per dimension (range features that define the same subgroups but have different descriptions and might be associated to same or different original features in the data), only a single representative was used in further construction of rages of higher dimensions.  Therefore the top ranked range features -- the selected subgroup descriptions -- are diverse, though the subgroups of  samples that they define have overlaps.
The remaining columns in Table~\ref{largest_dataset_range_ranking} report the confusion matrix statistics and quality metrics defined in Section~\ref{SS.FRA}. The selected ranges were sorted according to weighted precision defined in Section~\ref{SS.FRA}. One can see that many ranges with precision $1$ have been selected, which include single ranges and range triplets. Actually, all the selected ranges were of high quality. The validator chose the highest ranked range for splitting, in order to further root-cause the seven false negatives (value $7$ occurs in column $FN$ in the first row of Table~\ref{largest_dataset_range_ranking}). The second iteration was $\ra$ run on the dataset obtained by dropping all positive samples except the seven false negatives. With the ranges reported by the second $\ra$ run, it was possible to split the seven test failures into subgroups of four, two and one failures, respectively, which resulted in full root-causing of all seven failures.

\begin{table*}
\begin{tabular}{rrrrrrrrrrrrr}
\toprule
 Dim &       FN &   TN &     TP &   FP &      TPR &      PPV &        Lift &    WRAcc &  F1Score &  Accuracy \\
\midrule
         1 &     7 &  200 &  2619 &    0 &  0.9973 &  1.0000 &  1.0762 &    0.5328 &  0.9986 &    0.9975 \\
         3 &    11 &  200 &  2615 &    0 &  0.9958 &  1.0000 &  1.0762 &    0.5327 &  0.9979 &    0.9961 \\
         3 &    34 &  200 &  2592 &    0 &  0.9871 &  1.0000 &  1.0762 &    0.5325 &  0.9935 &    0.9880 \\
         1 &    35 &  200 &  2591 &    0 &  0.9867 &  1.0000 &  1.0762 &    0.5324 &  0.9933 &    0.9876 \\
         1 &    40 &  200 &  2586 &    0 &  0.9848 &  1.0000 &  1.0762 &    0.5324 &  0.9923 &    0.9858 \\
         3 &    51 &  200 &  2575 &    0 &  0.9806 &  1.0000 &  1.0762 &    0.5322 &  0.9902 &    0.9820 \\
         3 &    58 &  200 &  2568 &    0 &  0.9779 &  1.0000 &  1.0762 &    0.5322 &  0.9888 &    0.9795 \\
         3 &    59 &  200 &  2567 &    0 &  0.9775 &  1.0000 &  1.0762 &    0.5321 &  0.9886 &    0.9791 \\
         3 &    62 &  200 &  2564 &    0 &  0.9764 &  1.0000 &  1.0762 &    0.5321 &  0.9881 &    0.9781 \\
         3 &    64 &  200 &  2562 &    0 &  0.9756 &  1.0000 &  1.0762 &    0.5321 &  0.9876 &    0.9774 \\
         3 &    65 &  200 &  2561 &    0 &  0.9752 &  1.0000 &  1.0762 &    0.5321 &  0.9874 &    0.9770 \\
         3 &    72 &  200 &  2554 &    0 &  0.9726 &  1.0000 &  1.0762 &    0.5320 &  0.9861 &    0.9745 \\
         1 &    73 &  200 &  2553 &    0 &  0.9722 &  1.0000 &  1.0762 &    0.5320 &  0.9859 &    0.9742 \\
         3 &    77 &  200 &  2549 &    0 &  0.9707 &  1.0000 &  1.0762 &    0.5319 &  0.9851 &    0.9728 \\
         3 &    78 &  200 &  2548 &    0 &  0.9703 &  1.0000 &  1.0762 &    0.5319 &  0.9849 &    0.9724 \\
         1 &   110 &  200 &  2516 &    0 &  0.9581 &  1.0000 &  1.0762 &    0.5315 &  0.9786 &    0.9611 \\
         1 &   114 &  200 &  2512 &    0 &  0.9566 &  1.0000 &  1.0762 &    0.5315 &  0.9778 &    0.9597 \\
         3 &   115 &  200 &  2511 &    0 &  0.9562 &  1.0000 &  1.0762 &    0.5314 &  0.9776 &    0.9593 \\
         1 &   120 &  200 &  2506 &    0 &  0.9543 &  1.0000 &  1.0762 &    0.5314 &  0.9766 &    0.9575 \\
         3 &   122 &  200 &  2504 &    0 &  0.9535 &  1.0000 &  1.0762 &    0.5314 &  0.9762 &    0.9568 \\
         1 &   124 &  200 &  2502 &    0 &  0.9528 &  1.0000 &  1.0762 &    0.5313 &  0.9758 &    0.9561 \\
         3 &   149 &  200 &  2477 &    0 &  0.9433 &  1.0000 &  1.0762 &    0.5310 &  0.9708 &    0.9473 \\
         3 &   155 &  200 &  2471 &    0 &  0.9410 &  1.0000 &  1.0762 &    0.5309 &  0.9696 &    0.9452 \\
         3 &  2520 &  200 &   106 &    0 &  0.0404 &  1.0000 &  1.0762 &    0.5013 &  0.0777 &    0.1083 \\
         2 &   478 &  198 &  2148 &    2 &  0.8180 &  0.9991 &  1.0752 &    0.5266 &  0.8995 &    0.8301 \\
         2 &   478 &  196 &  2148 &    4 &  0.8180 &  0.9981 &  1.0742 &    0.5262 &  0.8991 &    0.8294 \\
         2 &   478 &  196 &  2148 &    4 &  0.8180 &  0.9981 &  1.0742 &    0.5262 &  0.8991 &    0.8294 \\
         2 &   478 &  196 &  2148 &    4 &  0.8180 &  0.9981 &  1.0742 &    0.5262 &  0.8991 &    0.8294 \\
         2 &   114 &  193 &  2512 &    7 &  0.9566 &  0.9972 &  1.0732 &    0.5303 &  0.9765 &    0.9572 \\
         2 &   122 &  192 &  2504 &    8 &  0.9535 &  0.9968 &  1.0727 &    0.5300 &  0.9747 &    0.9540 \\
         2 &   114 &  191 &  2512 &    9 &  0.9566 &  0.9964 &  1.0723 &    0.5300 &  0.9761 &    0.9565 \\
         3 &    34 &  190 &  2592 &   10 &  0.9871 &  0.9962 &  1.0720 &    0.5308 &  0.9916 &    0.9844 \\
         2 &   148 &  190 &  2478 &   10 &  0.9436 &  0.9960 &  1.0718 &    0.5294 &  0.9691 &    0.9441 \\
         2 &   154 &  190 &  2472 &   10 &  0.9414 &  0.9960 &  1.0718 &    0.5293 &  0.9679 &    0.9420 \\
         1 &   114 &  189 &  2512 &   11 &  0.9566 &  0.9956 &  1.0715 &   0.5296 &  0.9757 &    0.9558 \\
         2 &    34 &  170 &  2592 &   30 &  0.9871 &  0.9886 &  1.0638 &    0.5275 &  0.9878 &    0.9774 \\
         2 &    63 &  158 &  2563 &   42 &  0.9760 &  0.9839 &  1.0588 &    0.5252 &  0.9799 &    0.9628 \\
         2 &    34 &  127 &  2592 &   73 &  0.9871 &  0.9726 &  1.0467 &    0.5205 &  0.9798 &    0.9621 \\
         2 &    72 &  128 &  2554 &   72 &  0.9726 &  0.9726 &  1.0467 &    0.5201 &  0.9726 &    0.9490 \\
         3 &    47 &  117 &  2579 &   83 &  0.9821 &  0.9688 &  1.0426 &    0.5186 &  0.9754 &    0.9540 \\
         1 &    32 &  106 &  2594 &   94 &  0.9878 &  0.9650 &  1.0385 &    0.5170 &  0.9763 &    0.9554 \\
         2 &    47 &  105 &  2579 &   95 &  0.9821 &  0.9645 &  1.0379 &    0.5167 &  0.9732 &    0.9498 \\
         2 &    54 &   99 &  2572 &  101 &  0.9794 &  0.9622 &  1.0355 &    0.5156 &  0.9707 &    0.9452 \\
         2 &     6 &   84 &  2620 &  116 &  0.9977 &  0.9576 &  1.0305 &    0.5137 &  0.9772 &    0.9568 \\
         1 &     6 &   64 &  2620 &  136 &  0.9977 &  0.9507 &  1.0231 &    0.5104 &  0.9736 &    0.9498 \\
         2 &    51 &   63 &  2575 &  137 &  0.9806 &  0.9495 &  1.0218 &    0.5097 &  0.9648 &    0.9335 \\
         1 &    41 &   53 &  2585 &  147 &  0.9844 &  0.9462 &  1.0183 &    0.5082 &  0.9649 &    0.9335 \\
         1 &    49 &   49 &  2577 &  151 &  0.9813 &  0.9446 &  1.0166 &    0.5074 &  0.9626 &    0.9292 \\
         1 &     2 &   21 &  2624 &  179 &  0.9992 &  0.9361 &  1.0074 &    0.5034 &  0.9666 &    0.9360 \\
\bottomrule
\end{tabular}
\caption {Range Analysis results in first iteration of $\rct$ construction on $Data8$, with $\ra$ option values $\rpafeat = 15$ and $\rpamrmr = 12$, and with quality functions $WRacc$, $Lift$ and a weighted average of $TPR$ and $FNR$}
\label{largest_dataset_range_ranking}
\end{table*}

\subsection{Coverage with Precision 1 on PM Datasets}

In Table~\ref{pos_coverage_ppv_all_one} we present positive samples coverage information on all eight datasets in our benchmark suite. Table~$\ref{pos_coverage_ppv_all_one}.a$ reports coverage information on positive samples (i.e., test failures) with single range features only, with $\ra$. The first column lists the datasets, the columns $positives$ and $negatives$  report the count of positive and negative samples in the dataset, receptively, and they are included here for convenience. Columns $ppv\_min$ and $ppv\_max$ report the min and max values of the precisions (PPV) of the selected single ranges. To achieve a high level of diversity among the selected range tuples, in addition to the heuristics mentioned above, a heuristic method from~\cite{lavrac2004subgroup,van2012diverse} was used to generate a diverse set of single ranges that penalizes selection of range features that cover samples already covered by earlier selected range features. In particular, in experiments reported in  Table~\ref{pos_coverage_ppv_all_one}, the weight $w$ that controls the degree of penalizing coverage of a sample by $i$-th rule is computed as $w = \gamma^i$, where $0 \leq \gamma \leq 1$ was set to $0.4$, and then (weighted) coverage for a candidate rule is computed based on the weights~\cite{lavrac2004subgroup,van2012diverse}.  Note that when $\gamma = 1$ the heuristic has no effect, while $\gamma = 0$ corresponds to ignoring the previously covered samples (or dropping them from the dataset). Columns $cov\_min$ and $cov\_max$ report the min and max coverage of positives samples by the selected single ranges (not all samples need to be covered with the same number of rules). The number of samples not covered by any of the selected single ranges are reported in column $not\_cov$, and for convenience in column $all\_cov$ we report number of all covered samples by single ranges. Each of the coverage columns $cov\_min,  cov\_max,  not\_cov$ and $all\_cov$ actually report two numbers separated by $``/''$: on the left to $``/''$ we report the respective coverage information for all the selected single ranges while to the right we present coverage information for selected single ranges with precision $1$. Range features with precision $1$ clearly separate a subset of positive samples from the negative samples thus they can be considered as high quality root-causing hints. From the data reported in columns $cov\_min$ and  $cov\_max$ we see that for a single positive sample there are multiple selected single ranges with precision $1$, which means that there are multiple root-causing hints for these samples and each one of them can potentially point to the real root cause.

Table~\ref{pos_coverage_ppv_all_one}.b reports the same information as Table~\ref{pos_coverage_ppv_all_one}.a, except now it covers selected single ranges and range pairs, and Table~\ref{pos_coverage_ppv_all_one}.c reports the same information for all range features -- single ranges, pairs and triplets.  Table~\ref{pos_coverage_ppv_all_one}.d reports the same information obtained using pysubgroup $\sd$, where the $depth$ parameter was set to $3$, to generate range tuples up to triplets, and the $result\_set\_size$ parameter for each dataset was set to the number of generated triplets in Table~\ref{pos_coverage_ppv_all_one}.c for $\gamma = 1$. The latter choice is justified by the fact that pysubgroup $\sd$ does not support the $\gamma$-based heuristic and with $\gamma = 1$ this heuristic has no effect (in $\sd$ and in $\ra$). As it can be seen from Table~\ref{pos_coverage_ppv_all_one}.c, for datasets $1 - 3,7,8$ only very few samples remain not-covered by selected ranges with precision $1$, and they require an additional iteration of the $\rct$ procedure.

\begin{table*}
$$(\ref{pos_coverage_ppv_all_one}.a):\,\, Coverage\,\, with\,\, single\,\, ranges\,\, only$$
\begin{tabular}{lrrrrrrrrr}
\toprule
 data &  positives &  negatives &  ppv\_min &  ppv\_max & cov\_min & cov\_max & not\_cov & all\_cov  \\
\midrule
Data1 &        237 &        247 &   0.6286 &   1.0000 &         2/1 &        12/7 &         0/3 &     237/234   \\
Data2 &        237 &        246 &   0.5203 &   1.0000 &         3/3 &       13/10 &        0/37 &     237/200   \\
Data3 &        237 &        247 &   0.4917 &   1.0000 &         9/1 &        17/8 &         0/5 &     237/232   \\
Data4 &        226 &         44 &   0.8427 &   1.0000 &         7/1 &        17/1 &       0/209 &      226/17   \\
Data5 &        463 &        291 &   0.6316 &   1.0000 &         5/1 &        13/1 &       0/331 &     463/132   \\
Data6 &       2373 &        142 &   0.9519 &   0.9983 &         3/0 &        15/0 &      0/2373 &      2373/0   \\
Data7 &       2626 &        200 &   0.9348 &   1.0000 &         1/1 &        16/3 &        0/73 &   2626/2553   \\
Data8 &       2626 &        200 &   0.9361 &   1.0000 &         3/1 &        16/9 &         0/2 &   2626/2624   \\
\bottomrule
\end{tabular}

$$(\ref{pos_coverage_ppv_all_one}.b):\,\,Coverage\,\, with\,\, single\,\, ranges\,\, and\,\, pairs$$
\begin{tabular}{lrrrrrrrrr}
\toprule
 data &  positives &  negatives &  ppv\_min &  ppv\_max & cov\_min & cov\_max & not\_cov & all\_cov  \\
\midrule
Data1 &        237 &        247 &   0.6286 &      1.0 &         4/1 &       22/13 &         0/1 &     237/236   \\
Data2 &        237 &        246 &   0.5203 &      1.0 &         7/1 &       21/11 &         0/5 &     237/232  \\
Data3 &        237 &        247 &   0.4917 &      1.0 &        18/1 &        30/8 &         0/5 &     237/232   \\
Data4 &        226 &         44 &   0.8427 &      1.0 &        18/1 &        47/3 &       0/158 &      226/68   \\
Data5 &        463 &        291 &   0.6316 &      1.0 &         7/1 &        22/8 &       0/163 &     463/300   \\
Data6 &       2373 &        142 &   0.9519 &      1.0 &         4/2 &       35/14 &      0/1279 &   2373/1094   \\
Data7 &       2626 &        200 &   0.9348 &      1.0 &         1/1 &       32/12 &        0/11 &   2626/2615   \\
Data8 &       2626 &        200 &   0.9361 &      1.0 &         4/1 &        33/9 &         0/2 &   2626/2624   \\
\bottomrule
\end{tabular}

$$(\ref{pos_coverage_ppv_all_one}.c):\,\,Coverage\,\, with\,\, single\,\, ranges,\,\, pairs\,\, and\,\, triplets$$
\begin{tabular}{lrrrrrrrrr}
\toprule
 data &  positives &  negatives &  ppv\_min &  ppv\_max & cov\_min & cov\_max & not\_cov & all\_cov   \\
\midrule
Data1 &        237 &        247 &   0.6286 &      1.0 &         5/1 &       37/25 &         0/1 &     237/236   \\
Data2 &        237 &        246 &   0.5203 &      1.0 &         7/1 &       38/23 &         0/5 &     237/232   \\
Data3 &        237 &        247 &   0.4917 &      1.0 &        24/2 &       54/20 &         0/5 &     237/232   \\
Data4 &        226 &         44 &   0.8427 &      1.0 &        22/1 &       78/12 &       0/146 &      226/80   \\
Data5 &        463 &        291 &   0.6316 &      1.0 &         9/1 &       32/13 &       0/150 &     463/313   \\
Data6 &       2373 &        142 &   0.9519 &      1.0 &         4/3 &       65/25 &      0/1279 &   2373/1094   \\
Data7 &       2626 &        200 &   0.9348 &      1.0 &         1/4 &       58/27 &        0/11 &   2626/2615   \\
Data8 &       2626 &        200 &   0.9361 &      1.0 &         4/2 &       60/34 &         0/2 &   2626/2624   \\
\bottomrule
\end{tabular}

$$(\ref{pos_coverage_ppv_all_one}.d):\,\,Coverage\,\, with\,\, single\,\, ranges,\,\, pairs\,\, and\,\, triplets\,\,using\,\,pysubgroup$$
\begin{tabular}{lrrrrrrrrr}
\toprule
 data &  positives &  negatives &  ppv\_min &  ppv\_max & cov\_min & cov\_max & not\_cov & all\_cov \\
\midrule
Data1 &        237 &        247 &   0.9831 &   1.0000 &         8/4 &        77/4 &         2/5 &     235/232  \\
Data2 &        237 &        246 &   1.0000 &   1.0000 &       57/57 &       57/57 &       37/37 &     200/200  \\
Data3 &        237 &        247 &   0.9590 &   1.0000 &        42/1 &        45/1 &         3/5 &     234/232  \\
Data4 &        226 &         44 &   0.9513 &   0.9887 &         5/0 &        95/0 &       1/226 &       225/0  \\
Data5 &        463 &        291 &   0.9976 &   1.0000 &        11/1 &       61/51 &       43/43 &     420/420  \\
Data6 &       2373 &        142 &   0.9983 &   0.9983 &        81/0 &        82/0 &      4/2373 &      2369/0  \\
Data7 &       2626 &        200 &   1.0000 &   1.0000 &       66/66 &       66/66 &       73/73 &   2553/2553  \\
Data8 &       2626 &        200 &   1.0000 &   1.0000 &       58/58 &       63/63 &         7/7 &   2619/2619  \\
\bottomrule
\end{tabular}
\caption {Positive samples coverage with PPV = all / PPV = 1, with $\gamma=0.4$, with top $15$ range tuples selected per dimension by the target procedure in $\ra$; and with pysubgroup $\sd$ (where the value of $\gamma$ has no effect)}
\label{pos_coverage_ppv_all_one}
\end{table*}

In Table~\ref{pos_coverage_gamma} we evaluate in more detail the effect of the selected range feature diversification heuristic from~\cite{lavrac2004subgroup,van2012diverse}, discussed above.  In particular, we evaluate positive samples coverage for values $\gamma = 1.0, 0.9, 0.8, \ldots, 0.0$, for $\ra$ runs with $\rpafeat = 15$ and $\rpafeat = 25$, in Tables~\ref{pos_coverage_gamma}.a and~\ref{pos_coverage_gamma}.b, respectively; and with pysubgroup $\sd$, in the last column called $PSG$. Unlike Table~\ref{pos_coverage_ppv_all_one}, we report the count of positive samples not covered by selected range tuples with precision $1$, since the selected range tuples with high precision (of dimensions up to $3$) cover all positive samples at least once for all values of $\gamma$, in all datasets, even if this diversification heuristic is not used, which corresponds to case $\gamma = 1$. While for datasets $3, 7, 8$ the value of $\gamma$ has no effect in terms of improving positive samples coverage, for other datasets the effect is pretty significant, especially for dataset $6$, where activating the heuristic reduces the number not covered positive samples very significantly. The row named $Total$ reports the sum of not-covered positive samples in all the eight datasets, while the row named $Wins$ reports for how many datasets the given column, that is, given value of $\gamma$, was among the winning values of $\gamma$ in minimizing the number of not covered positive samples.  As it can be observed, the value $\gamma = 0.4$ is best performing on $\ra$ runs with $\rpafeat = 15$ in terms of the number of wins and minimizing the number of not covered positive samples, and $\gamma = 0.4$ is also among of the top winning choices for $\ra$ runs with $\rpafeat = 25$. Details of $\ra$ runs with this winning value of $\gamma = 0.4$ were given in Table~\ref{pos_coverage_ppv_all_one}.

\begin{table*}
$$(\ref{pos_coverage_gamma}.a):\,\, RA\,\, runs\,\, with \,\, \rpafeat = 15 \,\, and \,\, \rpamrmr = 12$$
\begin{tabular}{lrrrrrrrrrrrr}
\toprule
 data &  1.0 &   0.9 &   0.8 &   0.7 &   0.6 &   0.5 &   0.4 &   0.3 &   0.2 &   0.1 &   0.0 & PSG \\
\midrule
Data1 &    2 &    1 &    1 &    3 &    1 &    1 &    1 &    1 &    8 &    3 &    3  & 5 \\
Data2 &    6 &    6 &    6 &    6 &    5 &    5 &    5 &    5 &    5 &    6 &    6  & 37 \\
Data3 &    5 &    5 &    5 &    5 &    5 &    5 &    5 &    5 &    5 &    5 &    5  & 5 \\
Data4 &  147 &  140 &  135 &  155 &  142 &  155 &  146 &  158 &  158 &  158 &  150 & 226\\
Data5 &  159 &  165 &  158 &  170 &  158 &  150 &  150 &  150 &  163 &  163 &  256 & 43 \\
Data6 & 1437 & 1279 & 1279 & 1279 & 1279 & 1279 & 1279 & 1279 & 1279 & 1279 & 1279 & 2373 \\
Data7 &   11 &   11 &   11 &   11 &   11 &   11 &   11 &   11 &   11 &   11 &   11 & 73 \\
Data8 &    2 &    2 &    2 &    2 &    2 &    2 &    2 &    2 &    2 &    2 &    2 & 7 \\
\midrule
Total & 1769 & 1609 & 1597 & 1631 & 1603 & 1608 & 1599 & 1611 & 1631 & 1627 & 1712 & 2769 \\
Wins  & 3       &  5     &  6     & 4       &  6      &  7      &   7    &  7      &  5     &  4      &  4      & 1 \\
\bottomrule
\end{tabular}

$$(\ref{pos_coverage_gamma}.b):\,\,RA\,\, runs\,\, with \,\, \rpafeat = 25 \,\, and \,\, \rpamrmr = 12$$
\begin{tabular}{lrrrrrrrrrrrr}
\toprule
 data &   1.0 &   0.9 &   0.8 &   0.7 &   0.6 &   0.5 &   0.4 &   0.3 &   0.2 &   0.1 &   0.0 & PSG\\
\midrule
Data1 &    1 &    3 &    3 &    3 &    3 &    3 &    1 &    1 &    1 &    3 &    3  & 5  \\
Data2 &    6 &    6 &    6 &    6 &    5 &    5 &    5 &    5 &    5 &    6 &    6  & 37 \\
Data3 &    3 &    1 &    1 &    3 &    3 &    3 &    3 &    3 &    3 &    3 &    3   & 5 \\
Data4 &  147 &  130 &  140 &  150 &  136 &  148 &  148 &  134 &  129 &  140 &  140 & 226 \\
Data5 &  232 &  150 &  148 &  145 &  143 &  143 &  142 &  142 &  148 &  148 &  237 & 43  \\
Data6 & 1337 & 1177 & 1177 & 1177 & 1177 & 1177 & 1177 & 1177 & 1177 & 1177 & 1178 & 2373 \\
Data7 &   11 &   11 &   11 &   11 &   11 &   11 &   11 &   11 &   11 &   11 &  11  & 73 \\
Data8 &    2 &    2 &    2 &    2 &    2 &    2 &    2 &    2 &    2 &    2 &    2 & 7 \\
\midrule
Total & 1739 & 1480 & 1488 & 1497 & 1480 & 1492 & 1489 & 1475 & 1476 & 1490 & 1580  & 2769 \\
Wins  & 3       &  4          &  4         & 3         &  4        &  4        &   6        &  6         &  6        &  3     &  2  &  1   \\
\bottomrule
\end{tabular}
\caption {Positive samples coverage with PPV = 1 for different values of $\gamma$, with $\rpafeat = 15$ and $\rpafeat = 25$ in the top and bottom tables, respectively, and with $\rpamrmr = 12$, with $\ra$ algorithm; and with pysubgroup $\sd$, with the same count of selected range tuples as with $\ra$ and $\gamma = 1$}
\label{pos_coverage_gamma}
\end{table*}

\subsection{Importance of Types of Engineered Features on PM Datasets}

Finally, we evaluate which types of features engineered from Power Management regression logs are selected most frequently. In Table~\ref{hist_features_ratio} we report the types of features that occur among the top $6$ ranked selected range tuples, for $\ra$ runs with $\rpafeat = 15$, $\rpamrmr = 12$, and $\gamma= 0.4$. These tables demonstrate that the count and timing features, for which we generate histogram statistics as additional features for root causing, are selected very often; and the same is true for the histogram features generated from them. The occurrence features are also selected often, while the ordering features have the lowest representation among the top ranked range features.

\begin{table*}
\begin{tabular}{lrrrrrrrr}
\toprule
 data &                               features &   FN &  TN &   TP &  FP &     PPV \\
\midrule
Data1 &                 var367\_\_COUNT\_hist\_min &  104 & 247 &  133 &   0  & 1.0000  \\
Data1 &              var297\_\_COUNT\_\_0\_TO\_13547 &  119 & 247 &  118 &   0  & 1.0000  \\
Data1 &                 var14\_\_TIMING\_hist\_min &  129 & 247 &  108 &   0  & 1.0000  \\
Data1 &            var12\_\_TIMING\_hist\_skewness &  161 & 247 &   76 &   0  & 1.0000  \\
Data1 &                  var239\_\_COUNT\_\_0\_TO\_0 &  177 & 247 &   60 &   0  & 1.0000  \\
Data1 &                  var320\_\_COUNT\_\_0\_TO\_0 &  189 & 247 &   48 &   0  & 1.0000  \\
\midrule
Data2 &            var308\_\_COUNT\_\_1025\_TO\_1025 &   37 & 246 &  200 &   0  & 1.0000  \\
Data2 &                 var242\_\_COUNT\_hist\_max &   42 & 246 &  195 &   0  & 1.0000  \\
Data2 &              var412\_\_COUNT\_\_128\_TO\_128 &   84 & 246 &  153 &   0  & 1.0000  \\
Data2 &                 var14\_\_TIMING\_hist\_min &  129 & 246 &  108 &   0  & 1.0000  \\
Data2 &                 var452\_\_COUNT\_hist\_max &  189 & 246 &   48 &   0  & 1.0000  \\
Data2 &                 var380\_\_COUNT\_hist\_max &  190 & 246 &   47 &   0  & 1.0000  \\
\midrule
Data3 &               var6\_\_OCCURS\_\_167\_TO\_167 &   37 & 247 &  200 &   0  & 1.0000  \\
Data3 &                var9\_\_OCCURS\_\_0\_TO\_4092 &   63 & 247 &  174 &   0  & 1.0000  \\
Data3 &           var10\_\_TIMING\_\_6703\_TO\_13513 &  113 & 247 &  124 &   0  & 1.0000  \\
Data3 &                   var6\_\_OCCURS\_\_7\_TO\_7 &  122 & 247 &  115 &   0  & 1.0000  \\
Data3 &                  var56\_\_OCCURS\_\_0\_TO\_0 &  124 & 247 &  113 &   0  & 1.0000  \\
Data3 &                 var13\_\_TIMING\_hist\_min &  129 & 247 &  108 &   0  & 1.0000  \\
\midrule
Data4 &            var14\_\_TIMING\_hist\_kurtosis &  209 &  44 &   17 &   0  & 1.0000  \\
Data4 &        var45\_\_OCCURS\_\_116609\_TO\_358880 &  182 &  44 &   44 &   0  & 1.0000  \\
Data4 &                 var60\_\_TIMING\_hist\_max &  201 &  44 &   25 &   0  & 1.0000  \\
Data4 &                 var115\_\_OCCURS\_\_5\_TO\_5 &  184 &  44 &   42 &   0  & 1.0000  \\
Data4 &                  var46\_\_OCCURS\_\_1\_TO\_8 &  183 &  43 &   43 &   1  & 0.9773  \\
Data4 &               var8\_\_OCCURS\_\_167\_TO\_175 &  183 &  43 &   43 &   1  & 0.9773  \\
\midrule
Data5 &            var14\_\_TIMING\_hist\_kurtosis &  331 & 291 &  132 &   0  & 1.0000  \\
Data5 &              var11\_\_OCCURS\_\_0\_TO\_11750 &  339 & 291 &  124 &   0  & 1.0000  \\
Data5 &         var103\_\_OCCURS\_\_28455\_TO\_62818 &  346 & 291 &  117 &   0  & 1.0000  \\
Data5 &               var8\_\_OCCURS\_\_167\_TO\_175 &  373 & 291 &   90 &   0  & 1.0000  \\
Data5 &                 var14\_\_TIMING\_hist\_max &  384 & 291 &   79 &   0 &  1.0000  \\
Data5 &            var16\_\_OCCURS\_\_6144\_TO\_6144 &  403 & 291 &   60 &   0  & 1.0000  \\
\midrule
Data6 &                                                                   var3 & 1606 & 142 &  767 &   0  & 1.0000  \\
Data6 & var157\_\_OCCURS\_\_318767104\_TO\_318767104 & 1608 & 142 &  765 &   0  & 1.0000  \\
Data6 &                  var10\_\_OCCURS\_\_0\_TO\_0 & 1616 & 142 &  757 &   0  & 1.0000  \\
Data6 & var188\_\_OCCURS\_\_423624706\_TO\_427819011 & 1636 & 142 &  737 &   0  & 1.0000 \\
Data6 &                   var0\_\_OCCURS\_\_4\_TO\_4 & 1747 & 142 &  626 &   0  & 1.0000  \\
Data6 &                  var84\_\_OCCURS\_\_0\_TO\_0 &   28 & 138 & 2345 &   4 & 0.9983  \\
\midrule
Data7 &                                   var3 &   73 & 200 & 2553 &   0  & 1.0000  \\
Data7 &                                 var166\_\_ORDERING &  483 & 200 & 2143 &   0  & 1.0000  \\
Data7 &                                   var5 &  948 & 200 & 1678 &   0  & 1.0000  \\
Data7 &               var6\_\_OCCURS\_\_167\_TO\_167 &  355 & 200 & 2271 &   0  & 1.0000  \\
Data7 & var159\_\_OCCURS\_\_318767104\_TO\_318767104 &  369 & 200 & 2257 &   0  & 1.0000  \\
Data7 &                   var6\_\_OCCURS\_\_7\_TO\_7 &  374 & 200 & 2252 &   0 &  1.0000  \\
\midrule
Data8 &               var6\_\_OCCURS\_\_167\_TO\_167 &    7 & 200 & 2619 &   0  & 1.0000  \\
Data8 & var159\_\_OCCURS\_\_318767104\_TO\_318767104 &   35 & 200 & 2591 &   0  & 1.0000  \\
Data8 &                   var6\_\_OCCURS\_\_7\_TO\_7 &   40 & 200 & 2586 &   0  & 1.0000  \\
Data8 &                                   var3 &   73 & 200 & 2553 &   0  & 1.0000  \\
Data8 &                  var86\_\_OCCURS\_\_0\_TO\_0 &  110 & 200 & 2516 &   0  & 1.0000  \\
Data8 &                  var12\_\_OCCURS\_\_0\_TO\_0 &  120 & 200 & 2506 &   0  & 1.0000 \\
\bottomrule
\end{tabular}
\caption {Types of selected top ranked range tuples for RA runs with $\rpafeat = 15$, $\rpamrmr = 12$ and $\gamma = 0.4$}
\label{hist_features_ratio} 
\end{table*}

\section{Related Work and Discussion}\label{S.related}

\subsection{Subjective Measures of Interestingness of Subgroups}

Work~\cite{lavravc2004decision} discusses three use cases in other application domains -- one medical, two marketing -- with the aim to highlight advantages of $\sd$ methodology and the wide range of algorithms and quality functions developed under $\sd$ for the task of identifying relevant subgroups of a population.  While the above application domains are pretty remote from validation, big part of discussion and learning from~\cite{lavravc2004decision} apply to our work as well, and we repeat here some of the most relevant arguments (see Sections 3 and 8, ~\cite{lavravc2004decision}).

One can distinguish between \emph{objective quality measures} and \emph{subjective measures of interestingness}~\cite{silberschatz1995subjective}. Both the objective and subjective measures need to be considered in order to solve subgroup discovery tasks. Which of the quality criteria are most appropriate depends on the application. Obviously, for automated rule induction it is only the objective quality criteria that apply. However, for evaluating the quality of induced subgroup descriptions and their usefulness for decision support, the subjective criteria are more important, but also harder to evaluate.
Below is a list of subjective measures of interestingness:
\begin{enumerate}
\item \emph{Usefulness.} Usefulness is an aspect of rule interestingness which relates a finding to the goals of the user~\cite{klosgen1996multipattern}.
\item \emph{Actionability.} A rule (pattern) is interesting if the user can do something with it to his or her advantage~\cite{piatetsky1994interestingness,silberschatz1995subjective}.
\item \emph{Operationality.} Operationality is a special case of actionability. Operational knowledge enables performing an action which can operate on the target population and change the rule coverage. It is therefore the most valuable form of induced knowledge~\cite{lavravc2004decision}.
\item \emph{Unexpectedness.} A rule (pattern) is interesting if it is surprising to the user~\cite{silberschatz1995subjective}.
\item \emph{Novelty.} A finding is interesting if it deviates from prior knowledge of the user~\cite{klosgen1996multipattern}.
\item \emph{Redundancy.} Redundancy amounts to the similarity of a finding with respect to other
findings; it measures to what degree a finding follows from another one~\cite{klosgen1996multipattern}, or to what degree multiple findings support the same claims.
\end{enumerate}

The main reason why our root-causing approach is not fully automated is that we rely on validator's feedback to decide when splitting of failure subgroups is not required any further.  For example, to decide whether all failures in each subgroup of failures at the leaves of the $\rct$ constructed so far \emph{have the same root cause, with actionable insights}. Even if the subgroup descriptions -- the range tuples -- were generated automatically so that all subgroup descriptions together cover all the positive samples and only positive ones, and the samples covered by these subgroup descriptions are disjoint, it is not guaranteed that these subgroups are the right ones, and that the descriptions of these subgroups are the right ones.\footnote{There may be many such sets of subgroup descriptions in general, and there might be none, say in case the dataset has a sample that is both positive and negative; in the latter case, adding more features to the dataset might help to resolve such conflicts in the data and allow for cleaner separation of subgroups of positive samples from negative ones using range tuples or rules.} Indeed, as it can be seen from the experimental evaluation is Section~\ref{S.results}, a failing test might have multiple explanations through rules with precision~$1$, and not all of them need to point to the right root cause. Formal specifications enabling automating such a decision (to keep human out of the loop / eliminate the need for an oracle) do not exist, and as discussed in the introduction, such decisions can also depend on the context and goals of root-causing, and on how deep root-causing should go. We expect that further automation can be added in this direction by mining expert validator's \emph{decisions} through labeling the identified intermediate root-causing hints and the rules that can isolate the associated subgroups of failures, by extending ideas proposed in Subsection~\ref{SS.mining}. These rules can be reused for predicting and classifying failure root causes in future regressions. In particular, rules and models learned during pre-silicon validation can be reused for post-silicon validation and later for tracking anomalous behaviors during in-field operation.

\subsection{Comparing $\sd$ with $\rl$ and Decision Trees for Root-causing}

Work~\cite{lavravc2004decision} also compares $\sd$ methodology with decision trees and classification rule learning. Here are the main relevant points:

\begin{enumerate}
\item The usual goal of classification rule learning is to generate separate models, one for each class, inducing class characteristics in terms of properties (features) occurring in the descriptions of training examples. Therefore, classification rule learning results in \emph{characteristic} descriptions, generated separately for each class by repeatedly applying the \emph{covering algorithm}, meaning that the positive samples covered by already generated rules are dropped from the dataset when generating a new rule. 
\item In decision tree learning, on the other hand, the rules which can be formed from paths leading from the root node to class labels in the leaves represent \emph{discriminant} descriptions, formed from properties that best discriminate between the classes. As rules formed from decision tree paths form discriminant descriptions, they are inappropriate for solving subgroup discovery tasks which aim at describing subgroups by their characteristic properties.
\item Subgroup discovery is seen as classification rule learning  by treating subgroup $cond$ as a rule $cond \longrightarrow class$, where usually $class$ is the positive class. However, the fact that in classification rule learning the rules have been generated by a covering algorithm hinders their usefulness for subgroup discovery. Only the first few rules induced by a
covering algorithm may be of interest as subgroup descriptions with sufficient coverage. Subsequent rules are induced from smaller and strongly biased example subsets, excluding the positive examples covered by previously induced rules. This bias prevents the covering algorithm from inducing descriptions uncovering significant subgroup properties of the entire population.
\end{enumerate}

In the context of root-causing from regression test trace logs, the above statements from~\cite{lavravc2004decision} mean that both decision tree and rule-based algorithms are suitable for separating the two classes (fail vs pass), which is the explicit aim of the decision tree algorithms, but this is not enough since we want to also separate between subgroups of failures as not all failures have the same root cause; and here decision tree algorithms with discriminating description are less appropriate than subgroup discovery techniques with characteristic descriptions which aim at finding subgroups of failures with the same explanation (same root cause).

Our root-causing task is different from both $\sd$ and $\rl$ tasks in that the dataset that we work with can change during the root-causing procedure: we might add and/or remove some of the features in the initial input dataset, and might drop some of the positive samples in the initial dataset, in order to focus the root-causing procedure to one or more relevant modules and to a subgroup of positive samples. In addition, we want to learn subgroups of failures with the same root cause (which is a task relevant to $\sd$) as well as to learn rules that can serve as predictors of previously learned root causes in the future (which is a task relevant to $\rl$). Therefore, both $\sd$ and $\rl$ are relevant for our root-causing procedure, and so is $\ra$ which was designed to serve both above tasks. Both the objective and subjective measures of quality and interestingness discussed earlier remain relevant for this more general setting.

Work~\cite{chen2004failure} uses a modified version of $C4.5$ decision trees to root-cause failures in an internet service system, and demonstrates their usefulness. However, due to the \emph{classification} mindset in searching for root causes, which requires rules with high quality \emph{discriminant} descriptions, their tree splitting and branch selection heuristics prefer branches that maximize positive samples at the respective leaf nodes (see the \emph{ranking} criterion in Section 3.2), and operate under assumption that there are only a few independent sources of error (see the \emph{noise filtering} criterion in Section 3.2). Therefore, unlike the $\sd$ algorithms, their search algorithm does not encourage splitting subgroups of failing samples further even if the latter would result into higher quality subgroups according to relevant quality functions like \emph{WRAcc} or \emph{Lift}. As a consequence,  the approach in ~\cite{chen2004failure} does not fully adequately address the challenge of root-causing failures caused by a combination of multiple factors, which are usually most difficult to debug. Indeed, for the root-causing example discussed in Section~\ref{S.results} in some detail (for dataset $Data8$, with $\ra$ results in first $\rct$ iteration reported in Table~\ref{largest_dataset_range_ranking}), the final tree generated by the decision tree approach in~\cite{chen2004failure} would not contain branches leading to the three final subgroups with full root causes identified by our root-causing procedure, since each of these three subgroups cover fewer positive samples than the subgroup of the seven failures obtained by a first split that maximizes the number of positive samples in one of the branches.

\subsection{Comparing  $\sd$ with Other Approaches to Root-causing from Traces}

Work~\cite{mariani2008automated} proposed to learn behavior of the system from passing execution traces and compare failing runs against the learned models to identify the accepted and the unaccepted event (sub)sequences. The latter -- the suspicious sequences --  are presented to validator together with a representation of the behaviors that are acceptable instead of them, according to the learned models.
While the identified illegal event sequences localize well the bug locations within respective modules and serve as valuable input to the validators, further root-causing might be required since the learned models are abstract and do not capture all fine-grained legal behaviors (real systems are too complex to be fully learned as automata, not every event from the traces will be part of the learned model). 
The root-causing techniques proposed in our work can be considered complementary to the above research goal: we already have a well-defined specification for the flow of interest, thus it is straightforward to determine where the failing trace stops to conform to it, and instead our aim is to find observable event combinations (not necessarily occurring in the flow specification) that jointly explain the root cause. 

In~\cite{pal2021feature}, the authors address the root-causing problem in post-silicon SoC (System On Chip) validation. They apply \emph{clustering} to features engineered from traces of message-passing protocols at SoC level.\footnote{Their usage of term \emph{message} is similar to our usage of term \emph{event}, and messages might have additional attributes: ``In SoC designs, a message can be viewed as an assignment of Boolean values to the interface signals of a hardware IP''.} Their approach to feature engineering from traces is very different from ours. The idea that they follow is as follows: ''Logical bugs in designs can be considered as triggering \emph{corner-case} design behavior; which is \emph{infrequent} and \emph{deviant} from normal design behavior. In ML parlance, \emph{outlier detection} is a technique to identify \emph{infrequent} and \emph{deviant} data points, called \emph{outliers} whereas normal data points are called \emph{inliers}''. As a consequence, their task is to identify patterns of consecutive messages that cause the bugs, which they call \emph{anomalous message sequences}. The infrequency criterion can be captured through the entropy associated to message sequences and the deviancy is captured through a distance measure defined on the sequences. The techniques proposed in our work have been used as part of product development in the pre-silicon stage to root-cause regression suite failures, and at the early stages of product development often there are more failing tests than there are passing ones (this is the case for $5/8$ datasets used in our experiments). And there usually are many reasons for failures. Thus the outlier detection approach adopted in~\cite{pal2021feature} is more relevant at late stages of the product development and post-silicon when the bugs are relatively rare. 

To summarize, once trace logs are processed and a dataset is engineered, for which we proposed multiple widely applicable techniques of feature engineering, many ML and statistical approaches can be applied for the purpose of root-causing failures, and these techniques are often complementary. We believe that $\rl$ and $\sd$ based techniques are among the most suitable techniques for root-causing regression test failures when there might be multiple causes for failures and each cause for failure might require analyzing complex interplay between multiple factors to explain it. In addition, semi-supervised or unsupervised ML techniques like clustering, outlier direction, anomaly detection novelty detection, change detection and more can be used.

\section{Conclusions and Future Work}\label{S.future}

In this work we have proposed a new method for applying ML techniques on datasets engineered from a large set of test logs in order to generate powerful root-causing hints which can be mined and reused. The methodology proposed in this work has been applied successfully to debug $\pc$ flow of Intel's Power Management protocol.

While our focus was on root-causing regression test failures, and in particular on Power Management flows, our approach is very widely applicable. Indeed, the types of features that we generate from traces apply also for cases when there is no specification of the flow or when the flow is specified but its instances cannot be tracked accurately in the trace, and only pass and fail labels are known for each trace log. In that case, all events become non-flow events, and occurrence, count, ordering and timing features are defined for these events as well. Furthermore, the ability to encode sequential event occurrence information in trace logs through distributions and root-causing using properties of these distributions makes our approach very general. As already mentioned, when pass and fail labels are not available for trace logs, encoding of sequential event occurrence information into distributions still works, and unsupervised ML techniques can be used instead of $\RL$ and $\SD$ techniques.

As far as $\RL$ and $\SD$ are concerned, our main contribution is extension of these research frameworks  to one where the input dataset is not fixed and can be refined based on learning acquired through applications of $\rl$ and $\sd$ algorithms: both features and samples can be added to the input dataset or removed from it. As a consequence,  our framework supports much more realistic usages of $\rl$ and $\sd$. In addition,  introduction of quality orderings allows more freedom in specifying criteria for rule selection and we believe  they will be useful for other applications of $\rl$ and $\sd$ as well.

Learning sequential rules is somewhat similar to learning assertions from traces in the form of event sequences or automata, or in a richer formalism that can directly capture the concurrency information, but now the intention is to learn patterns of \emph{incorrect} scenarios rather than the correct ones (where incorrect scenarios are not necessary rare scenarios). In our iterative root-causing and learned bug mining approach, the rules can be mined by validation expert as sequential rules or in other formats. Further automation of mining sequential rules that capture bug patterns is an important direction for future work.

\section{Acknowledgments}

We would like to thank Yossef Lampe, Eli Singerman, Yael Abarbanel, Orly Cohen, Vladislav Keel and Dana Musmar for their contributions to this work: contributions to discussions,  implementation of respective parts of the root-causing tool, initial experiments, and actual usage in Intel products.

\bibliography{debug_archive_v2}{}

\begin{thebibliography}{10}

\bibitem{atzmueller2015subgroup}
Martin Atzmueller.
\newblock Subgroup discovery.
\newblock {\em Wiley Interdisciplinary Reviews: Data Mining and Knowledge
  Discovery}, 5(1):35--49, 2015.

\bibitem{billari2006timing}
Francesco~C Billari, Johannes F{\"u}rnkranz, and Alexia Prskawetz.
\newblock Timing, sequencing, and quantum of life course events: A machine
  learning approach.
\newblock {\em European Journal of Population/Revue Europ{\'e}enne de
  D{\'e}mographie}, 22:37--65, 2006.

\bibitem{chen2004failure}
Mike Chen, Alice~X Zheng, Jim Lloyd, Michael~I Jordan, and Eric Brewer.
\newblock Failure diagnosis using decision trees.
\newblock In {\em International Conference on Autonomic Computing, 2004.
  Proceedings.}, pages 36--43. IEEE, 2004.

\bibitem{chen2013simulation}
Wen Chen, Li-Chung Wang, Jay Bhadra, and Magdy Abadir.
\newblock Simulation knowledge extraction and reuse in constrained random
  processor verification.
\newblock In {\em Proceedings of the 50th Annual Design Automation Conference},
  pages 1--6, 2013.

\bibitem{clark1989cn2}
Peter Clark and Tim Niblett.
\newblock The cn2 induction algorithm.
\newblock {\em Machine learning}, 3(4):261--283, 1989.

\bibitem{de2013mrmre}
Nicolas De~Jay, Simon Papillon-Cavanagh, Catharina Olsen, Nehme El-Hachem,
  Gianluca Bontempi, and Benjamin Haibe-Kains.
\newblock mrmre: an r package for parallelized mrmr ensemble feature selection.
\newblock {\em Bioinformatics}, 29(18):2365--2368, 2013.

\bibitem{ding2005minimum}
Chris Ding and Hanchuan Peng.
\newblock Minimum redundancy feature selection from microarray gene expression
  data.
\newblock {\em Journal of bioinformatics and computational biology},
  3(02):185--205, 2005.

\bibitem{fraer2014visual}
Ranan Fraer, Doron Keren, Zurab Khasidashvili, Alexander Novakovsky, Avi Puder,
  Eli Singerman, Eran Talmor, Moshe~Y Vardi, and Jin Yang.
\newblock From visual to logical formalisms for soc validation.
\newblock In {\em 2014 Twelfth ACM/IEEE Conference on Formal Methods and Models
  for Codesign (MEMOCODE)}, pages 165--174. IEEE, 2014.

\bibitem{furnkranz2012foundations}
Johannes F{\"u}rnkranz, Dragan Gamberger, and Nada Lavra{\v{c}}.
\newblock {\em Foundations of rule learning}.
\newblock Springer Science \& Business Media, 2012.

\bibitem{guyon2003introduction}
Isabelle Guyon and Andr{\'e} Elisseeff.
\newblock An introduction to variable and feature selection.
\newblock {\em Journal of machine learning research}, 3(Mar):1157--1182, 2003.

\bibitem{hirsch2020root}
Thomas Hirsch and Birgit Hofer.
\newblock Root cause prediction based on bug reports.
\newblock In {\em 2020 IEEE International Symposium on Software Reliability
  Engineering Workshops (ISSREW)}, pages 171--176. IEEE, 2020.

\bibitem{hollander2001language}
Yoav Hollander, Matthew Morley, and Amos Noy.
\newblock The e language: A fresh separation of concerns.
\newblock In {\em Proceedings Technology of Object-Oriented Languages and
  Systems. TOOLS 38}, pages 41--50. IEEE, 2001.

\bibitem{IntelCore11thGEN}
{{Intel Corporation}}.
\newblock {\em 11th Generation Intel Core TM Processor Desktop, Datatsheet
  Volume 1}.
\newblock Intel Corporation, Santa Clara, California, United States, 2022.

\bibitem{katz2011learning}
Yoav Katz, Michal Rimon, Avi Ziv, and Gai Shaked.
\newblock Learning microarchitectural behaviors to improve stimuli generation
  quality.
\newblock In {\em Proceedings of the 48th Design Automation Conference}, pages
  848--853, 2011.

\bibitem{khasidashvili2019range}
Zurab Khasidashvili and Adam~J Norman.
\newblock Range analysis and applications to root causing.
\newblock In {\em 2019 IEEE International Conference on Data Science and
  Advanced Analytics (DSAA)}, pages 298--307. IEEE, 2019.

\bibitem{khasidashvili2021feature}
Zurab Khasidashvili and Adam~J Norman.
\newblock Feature range analysis.
\newblock {\em International Journal of Data Science and Analytics},
  11(3):195--219, 2021.

\bibitem{klosgen1996multipattern}
Willi Klosgen.
\newblock A multipattern and multistrategy discovery assistant.
\newblock {\em Advances in knowledge discovery and data mining}, pages
  249--271, 1996.

\bibitem{kotsiantis2006discretization}
Sotiris Kotsiantis and Dimitris Kanellopoulos.
\newblock Discretization techniques: A recent survey.
\newblock {\em GESTS International Transactions on Computer Science and
  Engineering}, 32(1):47--58, 2006.

\bibitem{lavravc2004decision}
Nada Lavra{\v{c}}, Bojan Cestnik, Dragan Gamberger, and Peter Flach.
\newblock Decision support through subgroup discovery: three case studies and
  the lessons learned.
\newblock {\em Machine Learning}, 57:115--143, 2004.

\bibitem{lavrac2004subgroup}
Nada Lavrac, Branko Kavsek, Peter Flach, and Ljupco Todorovski.
\newblock Subgroup discovery with cn2-sd.
\newblock {\em J. Mach. Learn. Res.}, 5(2):153--188, 2004.

\bibitem{lemmerich2018pysubgroup}
Florian Lemmerich and Martin Becker.
\newblock pysubgroup: Easy-to-use subgroup discovery in python.
\newblock In {\em Joint European Conference on Machine Learning and Knowledge
  Discovery in Databases}, pages 658--662, 2018.

\bibitem{mammo2016bugmd}
Biruk Mammo, Milind Furia, Valeria Bertacco, Scott Mahlke, and Daya~S Khudia.
\newblock Bugmd: Automatic mismatch diagnosis for bug triaging.
\newblock In {\em 2016 IEEE/ACM International Conference on Computer-Aided
  Design (ICCAD)}, pages 1--7. IEEE, 2016.

\bibitem{manukovsky2020machine}
A~Manukovsky, Y~Shlepnev, Z~Khasidashvili, and E~Zalianski.
\newblock Machine learning applications for com based simulation of 112 gb
  systems.
\newblock {\em Signal Integrity Journal}, 2020.

\bibitem{manukovsky2021machine}
Alex Manukovsky, Yuriy Shlepnev, and Zurab Khasidashvili.
\newblock Machine learning based design space exploration and applications to
  signal integrity analysis of 112gb serdes systems.
\newblock In {\em 2021 IEEE 71st Electronic Components and Technology
  Conference (ECTC)}, pages 1234--1245. IEEE, 2021.

\bibitem{mariani2008automated}
Leonardo Mariani and Fabrizio Pastore.
\newblock Automated identification of failure causes in system logs.
\newblock In {\em 2008 19th International Symposium on Software Reliability
  Engineering (ISSRE)}, pages 117--126. IEEE, 2008.

\bibitem{pal2021feature}
Debjit Pal and Shobha Vasudevan.
\newblock Feature engineering for scalable application-level post-silicon
  debugging.
\newblock {\em arXiv preprint arXiv:2102.04554}, 2021.

\bibitem{piatetsky1994interestingness}
Gregory Piatetsky-Shapiro and Christopher~J Matheus.
\newblock The interestingness of deviations.
\newblock In {\em Proceedings of the AAAI-94 workshop on Knowledge Discovery in
  Databases}, volume~1, pages 25--36, 1994.

\bibitem{silberschatz1995subjective}
Abraham Silberschatz and Alexander Tuzhilin.
\newblock On subjective measures of interestingness in knowledge discovery.
\newblock In {\em KDD}, volume~95, pages 275--281, 1995.

\bibitem{stearley2004towards}
John Stearley.
\newblock Towards informatic analysis of syslogs.
\newblock In {\em 2004 IEEE International Conference on Cluster Computing (IEEE
  Cat. No. 04EX935)}, pages 309--318. IEEE, 2004.

\bibitem{van2012diverse}
Matthijs Van~Leeuwen and Arno Knobbe.
\newblock Diverse subgroup set discovery.
\newblock {\em Data Mining and Knowledge Discovery}, 25:208--242, 2012.

\bibitem{wrobel1997algorithm}
Stefan Wrobel.
\newblock An algorithm for multi-relational discovery of subgroups.
\newblock In {\em Principles of Data Mining and Knowledge Discovery: First
  European Symposium, PKDD'97 Trondheim, Norway, June 24--27, 1997 Proceedings
  1}, pages 78--87. Springer, 1997.

\end{thebibliography}


\end{document}